\documentclass[final,authoryear,5p,times,twocolumn]{elsarticle}
\usepackage{newtxtext}
\usepackage[T1]{fontenc}
\usepackage{ae,aecompl}

\usepackage{graphicx}	
\usepackage{amsmath}	
\usepackage{amssymb}	
\usepackage{graphicx}
\usepackage{natbib}
\usepackage{array}
\usepackage{graphics}
\usepackage{latexsym}
\usepackage{amssymb}
\usepackage{amsmath}
\usepackage{fancyhdr}
\usepackage{morefloats}
\usepackage{bm}
\usepackage{color}
\usepackage{soul}

\newcommand\apjl{\@eapj@ApJLetters }

\title{On the   pressure equilibrium and timescales  in the scale free convection 
theory}

\author[r1,n3,r2]{S. Pasetto }  
\author[n2c]{C. Chiosi\corref{cor1}}  
\author[n3]{M. Cropper}  
\author[n2e]{E. Chiosi}   
\author[n4,n5]{D. Crnojevi\'c } 
\address[r1]{Dept. Integrated Mathematical Oncology (DIV Quantitative Science), H. Lee Moffitt Cancer Center and Research Institute
SRB-4, 12902 USF Magnolia Dr. Tampa (FL), 33612, USA, email: Stefano.Pasetto@Moffitt.org}  

\address[r2]{The Observatories of the Carnegie Institution for Science, 813 Santa Barbara St., Pasadena, CA 91101, USA}  
\address[n2c]{Department of Physics \& Astronomy, "G. Galilei", University of Padua, Vicolo 
dell'Osservatorio 2, Padua, Italy, email: cesare.chiosi@unipd.it }  
\address[n2e]{INAF-Astronomical Observatory of Padua, Vicolo 
dell'Osservatorio 5, Padua, Italy, email: echiosi@libero.it } 
\address[n3]{Mullard Space Science Laboratory, University College London, Holmbury St Mary, Dorking RH5 6NT, 
UK, email: m.cropper@ucl.ac.uk}   
\address[n4]{University of Tampa, 401 West Kennedy Boulevard, Tampa, FL 33606, USA, email: Denija.Crnojevic@ttu.edu }  
\address[n5]{Department of Physics \& Astronomy, Texas Tech University, Box 41051, Lubbock, TX 79409-1051, 
USA} 
\cortext[cor1]{Corresponding author}

\date{Accepted XXX. Received YYY; }

\begin{document}
\begin{frontmatter}
\begin{abstract}
Convection is  one of the fundamental energy transport processes in physics and  astrophysics, and 
its description is central to all stellar  models. 	
In the context of stellar astrophysics, the mixing length theory is the most successful approximation to 
handle the convection zones inside the stars because of its simplicity and rapidity. The price to pay is  
the mixing length parameter that is introduced to derive the velocity of convective elements, the temperature 
gradients in the convective regions and finally the energy flux carried by convection. The mixing length is a 
free parameter that needs to be calibrated on observational data.
Pasetto et al. (2014) have proposed a new theory that determines all the properties of convective 
regions and the convective transport of energy with no need for a  free parameter.
In this study, we aim to discuss the merits of this new approach and the limits of its applicability in 
comparison with the mixing length theory. 		
We present an analytical and numerical investigation of the main physical assumptions made by  Pasetto et 
al. (2014) and compare them with the counterparts of the mixing length theory.
We also present here the homogeneous isotropic limit of the Pasetto et al. (2014) theory and discuss some 
numerical 
examples to address and clarify misconceptions often associated with the new formalism. 	
\end{abstract}

\end{frontmatter}


\section{Introduction}\label{SecIntro}
Convection is one of the 
fundamental energy transport mechanisms in physics and  astrophysics, and 
its description is central to all stellar  models. Convection has been extensively studied 
over the years, and all attempts to model it have their 
origin in studies of turbulence, starting from the pioneering Kolmogorov work 
\citep[e.g.,][]{1941DoSSR..30..301K} up to the most recent computational simulations in physics 
\citep[e.g.,][]{2009AnRFM..41..165I, 
2010AnRFM..42..335L,2010JFM...646..527B,2010JFM...653..221B,2010LRSP....7....2R,2011AnRFM..43..219M,
2011A&A...528A.106S,2014AnRFM..46..567D,2014PNAS..11110961S} and astrophysics 
\citep[e.g.,][]{2014ApJ...785...90R,2015ApJ...798...51H,2015ApJ...808L..21C,2015ApJ...809...30A,
2015JFM...778..721G,2015ApJ...813...74J,2016ApJ...821L..17K,2016ApJ...832...71L,2017MNRAS.471..279C}. 
Excellent reviews can be found in \citet[][]{1998ApL&C..35..463F},  \citet[][]{2013imcp.book.....G} and 
\citet[][]{2017LRCA....3....1K}.

Stellar convection is customarily described by Mixing-Length Theory (MLT), a simplified analytical formulation 
of the energy transport
by convection, developed long ago by \citet{1958ZA.....46..108B}. The MLT stands on previous studies 
by \citet{1951ZA.....28..304B} and \citet{Prandtl}. In the literature there are many versions of the MLT, 
see e.g. \citet[][]{1968pss..book.....C} and \citet{1994sse..book.....K}, 
  or their modern revisits, e.g.
\citet{2004cgps.book.....W} and \citet[][]{2012sse..book.....K}, but all of them agree on the main lines.

The MLT stands on the basic (and largely justified by the common sense)  assumption that the convective 
elements during their existence and motion are in rigorous hydrostatic equilibrium with the surrounding 
medium. This is used to derive an elementary equation of motion for the convective elements under  
buoyancy and gravity forces, to evaluate their kinetic energy and velocity,  and then using the energy conservation 
principle to evaluate the total amount of energy stored by convective elements  when they come into 
existence, the 
energy lost by radiative processes into the medium, and finally the net amount of energy released to 
medium when the elements dissolve in it
\citep[see ][Chap 6, Sec.6.1]{2012sse..book.....K}. 
The MLT makes use of the
mixing length-scale to express the convective flux, velocity, and temperature gradients of the
convective elements and stellar medium. The mixing length-scale is taken to be proportional
to the local pressure scale-height, and the proportionality factor (the mixing-length parameter, MLP)
must be determined by comparing the stellar models to some calibrator, usually the Sun. No
strong arguments exist to suggest that the mixing-length parameter is the same in all stars
and at all evolutionary phases. 
All the stellar models in literature are affected by this basic drawback, i.e. the calibration of the MLP and the 
constancy of it with the stellar mass and evolutionary phases.

On one hand the assumption of rigorous hydrostatic equilibrium between convective elements and medium 
sounds reasonable at large scales because stars are in such a condition throughout their whole structure, 
including the convective zones in which the convective elements are supposed to originate, move and dissolve. 
On the other hand  by using this condition to derive the physical properties of convection we miss the 
correct derivation of the motion and the energy transport by convection. The MLT copes with this drawback by 
introducing the mixing length, which is more than a free adjustable parameter, and in reality puts the 
incomplete physics on the right track. Indeed each convective cell (which is the fundamental vector 
of the convection) is born, lives and always dies outside any form of equilibrium, i.e., always outside 
hydrostatic-equilibrium too. In other words in astrophysics, the detailed treatment of pressure in the 
convective zones is often omitted or hidden beyond the assumptions of the MLT. 

In a recent paper, \citet{2014MNRAS.445.3592P} (hereafter P14) presented an alternative theoretical framework 
that significantly differs from MLT in the fact that first it relaxes the local rigorous mechanical equilibrium 
between convective elements and medium, and second  it does not require 
any freely tuned parameter (i.e., a 
mixing length) to determine the energy transfer inside the stars. We will refer to this theory as to the 
scale-free convection theory (SFCT).

In this paper, we study the cardinal differences between these two theories with particular attention 
to their pressure treatment. We want to emphasize and clarify the pressure adjustment as addressed by both the 
SFCT and the MLT. This will cast light on the physical meaning of the mixing length parameter itself and on 
some misunderstanding generated by the assumption of the delayed and immediate pressure adjustment assumed by 
SFCT and MLT respectively.
In particular, we investigate the connection between the existence of a convective element, whose nature 
is by definition in a state of non-hydrostatic equilibrium, and the constant (time-independent) condition of 
hydrostatic equilibrium in which a star lives(\footnote{Thus, we are implicitly excluding stellar oscillations 
in this work for the sake of simplicity.}).

The plan of the paper is as follows. In Section \ref{pressure_field} we cast light on the basic issue about 
the pressure field surrounding convective elements i.e. whether these latter are always in pressure 
equilibrium with surrounding medium. In Section \ref{SecEssence} we elucidate the relevance of a correct 
temporal treatment of the 
pressure 
readjustment of the average convective element and its relation to the (always true) hydrostatic-equilibrium 
of the whole star, and with the aid of a simple semi-analytical case, we highlight the difference with 
respect 
to acceptance (not only) in the MLT that $Dp \equiv {p_e} - {p^\infty } = 0 \forall t\leq\infty$. In Section 
\ref{example}, with a simple numerical case,  we check the SFCT  and the limits of the so-called uniqueness 
theorem of P14. In section \ref{SecLimitsOfValidity} we investigate numerical and theoretical limitations of 
the SFCT. In section \ref{SecConclusion}, we highlight the relevance of the correct physical treatment of the 
dynamics in the convective layers of a star.
Finally in \ref{AppendixA} we brief note on the non-inertial linear response theory for convective elements 
in spherical coordinates, of which the SFCT is a particular case.

\section{The pressure field around convective elements}\label{pressure_field}
 In common with MLT, the SFCT considers  a star to consists of  many layers. Each layer, say a layer  $L$, 
 is defined in the SFCT and 
MLT by its average pressure $p$, density $\rho$ and temperature $T$ respecting the classic hydrostatic 
equilibrium condition. The role of the SFCT, or the MLT, is to insert the amount of energy that has to be 
carried by the average convective element moving up and down the layers before the it starts to 
travel. If the pressure readjustment of any convective element is instantaneous, as in the MLT, there is 
apparently no transportation problem. If the pressure readjustment of the average convective element is not 
instantaneous but delayed by a physically meaningful finite speed velocity, as in a time-dependent SFCT, we 
need to check that the element does not start before the energy is inserted and ready to be sent.

The vectors carrying the energy are the eddies/convective elements. An eddy is a blob of vorticity 
with its associated velocity field ${{\bm{v}}_0}$ inside a bounded stellar medium  
$L$ (\footnote{Mathematically, we can consider the layer to be  mapped with a reference frame 
${S_0}:{S_0}\left( {O,{\bm{x}};t} \right)$ with $L \subset {\mathbb{R}^4}$ assumed to be \textit{compact} 
(i.e., bounded and limited) with $O$ origin of the reference frame, spatial coordinates $\bm{x}$ and temporal 
coordinate $t$. We can refer to ${S_0}$ as to an inertial reference frame to leave the notation $S_1$ for the 
non-inertial reference frames.}).  In what follows,  we will refer  to any averaged convective element scheme 
as an eddy even though some of the adopted descriptions, e.g. the potential-flow approximation, do not 
consider vorticity in their formalism (they are curl-free). Any physical quantity in $L$ cannot extend to 
infinity, because the layer is finite, therefore not representative of the whole star, and it does not last 
forever. In $L$ the medium is described by  the Navier-Stokes equations and it is characterized by   an 
Equation of State (EoS) linking together  the state variables such as the pressure $p = p\left( {{\bm{x}};t} 
\right)$, density $\rho  = \rho \left( {{\bm{x}};t} \right)$, temperature $T = T\left( {{\bm{x}};t} \right)$ 
at any position $\bm{x}$ and  time $t$.

The pressure field surrounding an eddy is obtained from the theory of potential 
\citep[e.g.,][]{Kellogg,1975clel.book.....J}. This is before any assumption concerning the shapes of the 
eddies and the non-universal behavior of the convective turbulence. We start considering the fundamental 
proportionality relationship between pressure, density and the velocity field around a convective element 
\citep[e.g.,][Sec. 2]{1956RSPTA.248..369B} 

\begin{equation}\label{Eq03}
p\left( {{\bm{x}};t} \right) = \frac{\rho\left( {{\bm{x}};t} \right)}{{4\pi }}\nabla _{\bm{x}}^2 
{{{\left\| {\bm{x}} \right\|}^{ - 1}}} \int_{}^{} {{\bm{v}}_0^T\left( {{\bm{x'}};t} 
\right){{{\bm{v}}}_0}\left( {{\bm{x'}};t} \right)d{\bm{x'}}}  + ...,
\end{equation}
to the leading order, where  ${*^T}$ is the transpose
 element and ${\nabla ^n}\left[ * \right]$ the power-n gradient operator applied to what is immediately on 
its right. This is the fundamental equation of convective turbulence \citep[e.g.,][]{ 1956RSPTA.248..369B, 
1967JFM....27..581S}.

We remark on aspects of Eq.\eqref{Eq03}:
\begin{enumerate}
	\item Eq.\eqref{Eq03} is universally valid: it does not stand on the potential-flow approximation. It 
holds indeed for a fully rotational fluid too, where the concept of the eddy finds its natural definition.
	
	\item Eq.\eqref{Eq03} tells us how the pressure field associated with a convective element falls off 
and correlates with the motion of any other convective element across the generic layer, $L$, inside a star.
	
	\item The right-hand side (RHS) of Eq.\eqref{Eq03} is not constant, but retains an explicit 
dependence 
on time and position, meaning that the convective element is never required to be in equilibrium with its 
surrounding, and the pressure acting on it changes with time.
	
         \item Hence, if a convective cell when born is not in pressure equilibrium with the surrounding 
medium, the same bubble dies inside a star long before reaching the condition of hydrostatic pressure 
equilibrium with the medium too
(\footnote{Sec.\ref{SecFate},  \ref{AppendixA}, \citet{2015A&A...573A..48P}.}). If we define $Dp 
\equiv {p_e} - {p^\infty }$, then clearly holds:
	\begin{equation}\label{PressEq}
	Dp \equiv {p_e} - {p^\infty } \ne 0 \forall t<\infty
	\end{equation}
where ${p_e} \equiv p\left( {{{\bm{x}}_e};t} \right)$ is the pressure at a location $\bm{x_e}$ on the eddy 
surface and ${p^\infty } \equiv \mathop {\lim }\limits_{{\bm{x}} \to \partial L } p\left( {{\bm{x}};t} \right) 
< \infty$. This is always true if we assume $L$ to be compact by Weierstrass's theorem(\footnote{For our 
purposes, the Weierstrass theorems grants that a continuous function (e.g., the pressure $p$) on a compact set 
(e.g., $L$) is necessarily bounded.}) is the pressure ``far away'' from the eddy surface (e.g., at the 
topological boundary $\partial L$).
	
\end{enumerate}

\begin{figure*}
\centering{
	\includegraphics[height=80mm,width=160mm]{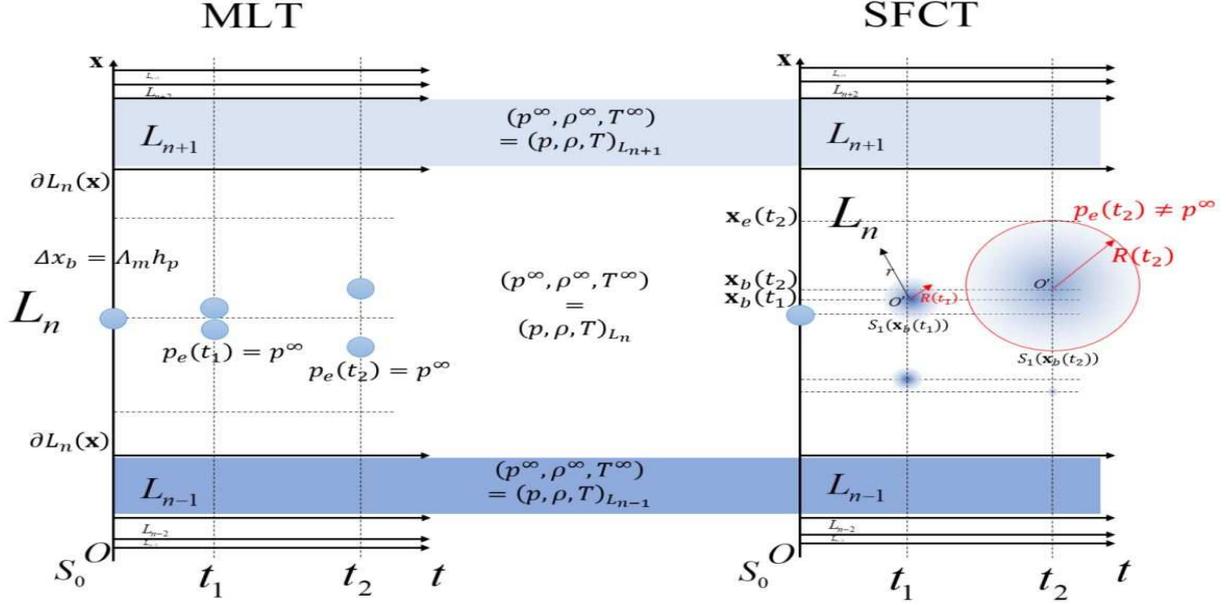}   }
	\caption{Pictorial representation of the temporal evolution of the same a convective element for MLT 
and SFCT. 
The star is discretized in $M$ layers in hydrostatic equilibrium, $L_n$, with $n=1,..,M$. The generic 
layer, ${{L}_{n}}$, is defined by the hydrostatic pressure, density, and temperature ${{\left( p,\rho ,T 
\right)}_{{{L}_{n}}}}$. For both MLT and SFCT, these values are assumed to be time-independent and far away 
from the convective element, i.e., for every layer $\left( {{p}^{\infty }},{{\rho }^{\infty }},{{T}^{\infty }} 
\right)={{\left( p,\rho ,T \right)}_{{{L}_{n}}}}\forall t$. For the SFCT, each convective element is never in 
hydrostatic equilibrium with its medium; and the pressure, temperature and density at the convective element 
surface are $\left\{ {{p}_{e}}\left( t \right),{{\rho }_{e}}\left( t \right),{{T}_{e}}\left( t \right) 
\right\} \ne \left\{ {{p}^{\infty }},{{\rho }^{\infty }},{{T}^{\infty }} \right\}$. Upper and lower border of 
the layer are a function of the position, $\partial {L_n} = \partial {L_n}\left( {\bm{x}} \right)$. The 
representation is purely indicative, the size of the layer $L_n$ is typically much larger than the size of the 
convective element, say $\left\| {{L_n}} \right\| \gg R\left( t \right)\forall t$ for $n = 2,..,M - 1$.}
	\label{FigSchemeTime}
\end{figure*}

Equation \ref{Eq03}  forms the basis of turbulence theory and  the transport of energy at different scales, times, 
and locations. This equation indeed shows how any small eddy can be considered as possible source (via the 
local pressure enhancement) of any other eddy in the environment under examination (i.e., $L$).

Probably the most straightforward eddy model in the literature is the one currently used in stellar 
astrophysics
 by the MLT (except for a few models of the Sun and other types of the star). Here an eddy is viewed as a 
(non-expanding) spherical body moving in an inertial reference frame ${S_0}\left( {O,{\bm{x}};t} \right)$.
While moving, the sphere instantaneously adjusts the pressure at its surface, i.e., the following condition 
is  always satisfied:
$$Dp = {p_e} - {p^\infty } = 0 \,\, \forall t.$$
The equation of motion for the barycenter, $x_b$, of the convective elements along its vertical motion is
$$\Delta {x_b} = \int_0^{{t_L}} {{{\dot x}_b}dt}  = {\Lambda _m}{h_p},$$
where ${\Lambda _m}$ is the Mixing-Length that is usually supposed to be proportional to the pressure scale 
length ${h_p}$, and $t_L$ is the lifetime of the convective element inside the layer $L$. This implies that 
while in the MLT the translational motion of the eddy is somehow taken into account, its expansion is ignored. 
In other words, if  $\left\{ {\bm{x}_b,\bm{x}_e} \right\}$ are the independent coordinates describing the two 
degrees of freedom of an eddy (i.e., its position $\bm{x}_b$, and size $\left\| {{{\bf{x}}_b} - {{\bf{x}}_e}} 
\right\|$, for the translation and expansion/contraction description, with $\bm{x}_e$ and element of the 
convective bubble surface), only one of them is taken into account. This is shown in Fig.\ref{FigSchemeTime}. 
The classical MLT is not entirely consistent with the temporal description of these two coordinates.
We understand that encoded in the free parameter ${\Lambda _m}$ there is not only the distance that a 
possible average element travels but much more:
\begin{enumerate}
	\item the average energy exchanged between the mean flow and turbulence;
	\item the differential behavior of intermittency (i.e., the irregular dissipation of kinetic energy) 
          at different layers of the star;
	\item the whole energy cascade as well as the temporal evolution required by the expansion in the 
         second ``omitted'' independent coordinate;
	\item and finally, that ${\Lambda _m}$ quantifies the amount of energy carried by convection in a star to 
         secure the observed luminosity and effective temperature.
\end{enumerate}

In the astrophysical context, a coherent  description of the  temporal evolution of the pressure and motion of 
a convective eddy in relation to its independent coordinates is present in the SFCT proposed by P14. This 
approach stems from a  more general theory of non-inertial gas linear-instabilities, where the Rayleigh-Taylor 
(RT) and Kelvin-Helmholtz (KH) instability treatment is included. As with the MLT, the SFCT assumes spherical 
symmetry for the convective eddy in order to simplify the mathematical formulation  in the context of the 
potential-flow formalism.  A description of the temporal and spatial pressure differences at the surface of  a 
convective element with respect to the medium has been investigated and presented by 
\citet{2016MNRAS.459.3182P}, hereafter P16, in the context of the SFCT. In that study (see their Fig.1)  the 
temporal evolution of the pressure at the surface is compared with the hydrostatic equilibrium pressure of the 
stellar layer under consideration. The description of the convective elements presented by P14 and P16 takes 
the physics of non-hydrostatic equilibrium of the average eddy into account already in the definition of a 
convective element and includes the dependence of it on time (unlike the case of the MLT).

\section{Capturing the essence of the SFCT}\label{SecEssence}
The correct description of the pressure and its variations essential for a correct formulation of the energy 
transport by convection/conduction/radiation, and the criteria for the onset of convection are indeed based on 
the comparison between the temperature over the pressure gradients $\nabla  \equiv \mathop {\lim 
}\limits_{{\bm{x}} \to \partial L } \frac{{\partial \ln T\left( {{\bm{x}};t} \right)}}{{\partial \ln p\left( 
{{\bm{x}};t} \right)}}$, and ${\nabla _e} \equiv \mathop {\lim }\limits_{{\bm{x}} \to {{\bm{x}}_e}} 
\frac{{\partial \ln T\left( {{\bm{x}};t} \right)}}{{\partial \ln p\left( {{\bm{x}};t} \right)}}$, etc. for 
every $t$.
Given the relevance of the pressure description inside a star, it is essential to highlight the implications 
of the pressure treatment adopted in the SFCT and to understand if the founding hypotheses of the SFCT can 
capture the real behavior of a convective blob rolling up in vortices of fluid.

\subsection{SFCT homogeneous-isotropic limit case}\label{SecHomoSFCT}
A fundamental assumption on which SFCT stands is that the expansion/contraction rate of a convective element 
is much larger than the motion of its barycenter. In Fig.\ref{FigSchemeTime} this is pictorially represented 
by the expansion of the convective element radius, $R$, being much larger than the vertical motion of the same 
element in the transition between two arbitrary instants $t_1$ and $t_2>t_1$.
If we follow the expansion/contraction in the non-inertial reference ${S_1}\left( {O',{\bm{r }}} \right)$ 
co-moving with the barycenter of the convective eddy at velocity ${v_{O'}}$ (relative to the inertial system 
$S_0(O; t)$ whose origin is at the center of the star $O$), we can write this assumption as $\left\| \bm{\dot 
r }_e \right\| = \dot R  \gg {v_{O'}}$.
We physically motivate this approach because eddies are supposed to die shortly after their birth and to 
dissolve by RT and KH instabilities thus releasing their energy to the surrounding medium (see, e.g, 
Sec.\ref{SecFate} and \ref{AppendixA}). If the expanding/ shrinking elements of the SFCT capture the 
essential behavior of physical eddies rolling up in turbulent vortices, then we should be able to achieve 
reasonably good results in the pressure description even if we neglect their vertical motion entirely. This 
means to investigate the behavior of the SFCT as it approaches the limit case of a homogeneous-isotropic 
medium, a situation that can be easily achieved by nullifying the acceleration, i.e., setting  
${{\bm{a}}_{O'}} = 0$, in the core equation of the SFCT(\footnote{This is identical to  Eq.(A.18) of 
\citet[][]{2015A&A...573A..48P} and a direct consequence of our Eq.\eqref{Eq03} from which it can be derived 
as shown in \citet{2014MNRAS.445.3592P}. }):

\begin{equation}\label{BernulliEq}
\frac{{\partial {\varphi _{\bm{v_1}}}}}{{\partial t}} + \frac{1}{2}\left\langle 
{{\nabla _{\bm{\xi}}}{\varphi _{\bm{v_1}}},{\nabla _{\bm{\xi}}}{\varphi _{\bm{v_1}}}} \right\rangle  + 
\frac{p}{\rho } = f\left( t \right) - \left\langle {{a_{O'}},{\bm{\xi}}} \right\rangle,
\end{equation}
with $\varphi_{\bm{v_1}}$ the velocity-potential flow for the stellar plasma in the non-inertial reference 
${S_1}\left( {O',{\bm{\xi }}} \right)$.

Without translation of the convective element the regime is that of homogeneous isotropic turbulence 
\citep[e.g.,][]{SagautP}. In this simplification there is no net convective flow: this would be the case in 
which convective elements carry all the energy transport with little vertical motion but with significant 
expansion, the opposite of what is supposed to occur with the MLT in which only the vertical motion is 
considered. The scientific case we are investigating should be considered as a mathematical idealization meant 
to capture the essence of the role played by the sole expansion as compared to the case of the sole vertical 
motions.

We describe the expansion of a convective element with the formalism of the velocity-potential flow  
\citep[e.g.,][]{1959flme.book.....L}. In the following, we will show that this simple description can capture 
the essence of the convective turbulence described by P14. In the absence of translation of the convective 
element, we can simplify our previous treatment in P14 from two reference frames moving relative to each other 
into a single  static one. It is convenient to rationalize the notation as follows: ${S_1}\left( {{O',\bm{\xi 
}}} \right) \equiv {S_0}\left( {{O,\bm{x}}} \right)\forall t$ hence, by the assuming spherical coordinates in 
${S_1}$, we write ${{S}_{0}}\left(O, \bm{x} \right) \equiv {{S}_{0}}\left(O, \bm{r} \right)={{S}_{0}}\left(O, 
r,\theta ,\phi  \right)$ where the system by spherical symmetry can be described merely by the evolution of 
the radius vector $\left\| {\bm{r}} \right\| \equiv r$ whose value at the convective element surface is the 
function of time ${\left\| {\bm{r}} \right\|_{\mathop{\rm e}\nolimits} }{\hat e_r} \equiv R(t)$.

\begin{figure}
	\includegraphics[width=\columnwidth]{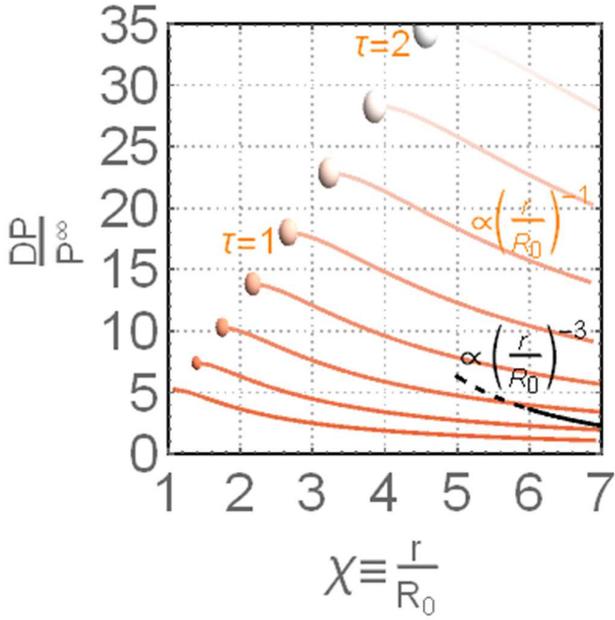}
	\caption{The temporal and spatial evolution of the relative pressure ration between the value at the 
bubble surface and the one at infinity. Normalized time runs from $\tau  = 0$ to $\tau  = 2$ from the bottom 
to the top. The lines start at the current surface of the spherical element position that is marked with a 
sphere whose radius increases according to Eq.\eqref{Eq09}. The profiles starting from the surface of the 
different spheres show the variation of the relative pressure difference as moving far away from a sphere 
(surface of the generic convective element) throughout large distances. The lines refer to the case of fluid 
in the irrotational potential-flow approximation. The dashed line shows (limited to one case),  the expected 
variation for a rotational flow. The disintegration of any generic convective element as time goes by is 
indicated here by drawing the spheres and lines with colors of lower and lower intensity.}
	\label{FigPressure}
\end{figure}

\begin{figure}
	\includegraphics[width=\columnwidth]{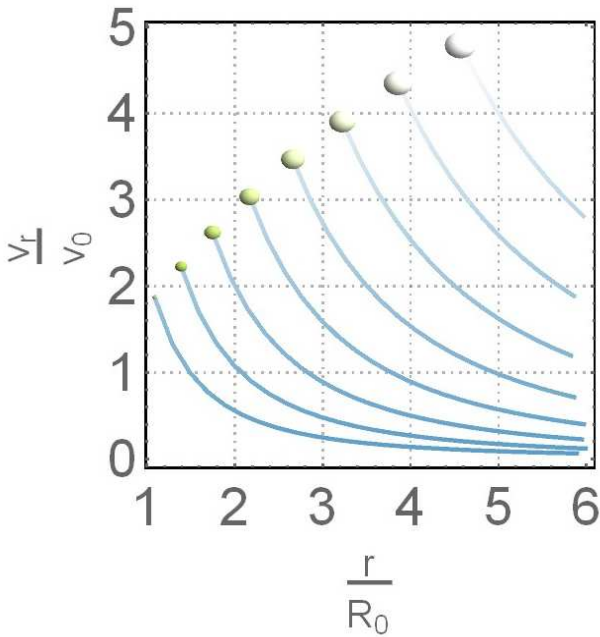}
	\caption{The same as in Fig.\ref{FigPressure} but for the velocity field surrounding the eddy.  }
	\label{FigVelocity}
\end{figure}

Under these assumptions,  the potential vector in $S_0$ is ${\Phi _{{{\bm{v}}_0}}} =  - 
\frac{{\dot R{R^2}}}{r}$ and the velocity field ${{\bm{v}}_0} = \frac{{\dot R{R^2}}}{{{r^2}}}{\hat e_r}$. 
Given these premises, at an arbitrary but fixed time $t$ the Bernoulli equation in ${{S}_{0}}$ becomes (see 
Eq.\eqref{BernulliEq} and Fig.\ref{FigSchemeTime}):
\begin{equation}\label{Eq04}
\frac{{\partial {\Phi _{{{\bm{v}}_0}}}}}{{\partial t}} + \frac{{{{\left\| {{{\bm{v}}_0}} \right\|}^2}}}{2} 
+ \frac{p}{\rho } = \text{cnst.},
\end{equation}
which with
\begin{equation}\label{Eq05}
\left\{ \begin{gathered}
\frac{{\partial {\Phi _{{{\bm{v}}_0}}}}}{{\partial t}} =  - \frac{{R\left( {2{{\dot R}^2} + R\ddot R} 
\right)}}{r} \hfill \\
\frac{{{{\left\| {{{\bm{v}}_0}} \right\|}^2}}}{2} = \frac{1}{2}\left\langle {\frac{{\partial {\Phi 
_{{{\bm{v}}_0}}}}}
{{\partial r}},\frac{{\partial {\Phi _{{{\bm{v}}_0}}}}}{{\partial r}}} \right\rangle  = 
\frac{1}{2}\frac{{{R^4}{{\dot R}^2}}}{{{r^4}}}, \hfill \\
\end{gathered}  \right.
\end{equation}
becomes:
\begin{equation}\label{Eq06}
\frac{{p\left( {r;t} \right)}}{{\rho \left( {r;t} \right)}} + \frac{1}{2}\frac{{{R^4}{{\dot R}^2}}}{{{r^4}}} 
- \frac{{R\left( {2{{\dot R}^2} + R\ddot R} \right)}}{r} = \text{cnst.}
\end{equation}
We set the constant by requiring that far away from the sphere we have $p\left( {\partial L;t} \right) = 
{p^\infty }$ 
and $\rho \left( {\partial L; t} \right) = \rho^\infty$ for all $t$ mainly because the size  $L$ of the layer 
is much larger than that of the eddy (see Fig.\ref{FigSchemeTime}). Note here that the formulation of the 
problem is entirely general, not requiring hydrostatic condition at the border of the domain of definition 
$\partial L$; on the contrary $p\left( {\partial L ;t} \right)$ could be taken to be any other pressure within 
$L$ and outside the eddy without invalidating these arguments. This is implicit to the sharing of pressure 
information introduced by  Eq.\eqref{Eq03} whose applicability is entirely general. Nevertheless, we will 
assume that the stellar convective layer is always in rigorous hydrostatic equilibrium as a star is, i.e., 
SFCT correctly assumes hydrostatic-equilibrium at ${\partial L}$ for each $L$ and each $t$. Therefore, with 
these considerations and relaxing the time dependence only in $\rho$, we can complete the physical description 
in Eq.\eqref{Eq06} with
\begin{equation}\label{Eq07}
\frac{{p\left( {r;t} \right)}}{\rho } + \frac{1}{2}\frac{{{R^4}{{\dot R}^2}}}{{{r^4}}} - 
\frac{{R\left( {2{{\dot R}^2} + R\ddot R} \right)}}{r} = \frac{{{p^\infty }}}{\rho^\infty },
\end{equation}
which simplifies as:
\begin{equation}\label{Eq08}	
\frac{{{{\dot R}^2}}}{2}\frac{{{R^4}}}{{{r^4}}} - \frac{R}{r}\left( {2{{\dot R}^2} + R\ddot R} \right) = 
\frac{{{p^\infty } - p\left( {r;t} \right)}}{\rho^\infty }.
\end{equation}
This is either a differential equation for $R\left( t \right)$ if the pressure is given or an equation for 
the pressure field if $R\left( t \right)$ is given.
We follow this second interpretation because from Eq.(B15) of P16 (see also Fig.A1 of P14) we have already 
learned that the interesting temporal evolution of the convective eddy is quadratic in time to the first 
order. This means that we can assume that the solutions of interest are those of type
\begin{equation}\label{Eq09}
\frac{R}{{{R_0}}} = 1 + \frac{{{t^2}}}{{t_0^2}},
\end{equation}
and we solve Eq.\eqref{Eq07} as an equation for the pressure field.
We look at a  generic sphere  instantaneously overlapping  the expanding eddy with 
$R \propto {t^2}$, set $r = R\left( t \right)$  in Eq.\eqref{Eq08}, and make use of Eq.\eqref{Eq09} to obtain:
\begin{equation}\label{Eq10}
\begin{gathered}
{\left. {\frac{{{{\dot R}^2}}}{2}\frac{{{R^4}}}{{{r^4}}} - \frac{R}{r}\left( {2{{\dot R}^2} + R\ddot R} 
\right)} \right|_{r = R}} = {\left. {\frac{{{p^\infty } - p\left( {r;t} \right)}}{\rho }} \right|_{p = {p_e}}} 
\Leftrightarrow,  \\
\frac{1}{2}\frac{{{\tau ^2}{{\left( {1 + \tau } \right)}^8}}}{{{\chi ^4}}} - \frac{{{{\left( {1 + \tau } 
\right)}^4}}}{\chi }\left( {\frac{1}{2} + \frac{{2{\tau ^2}}}{{{{\left( {1 + \tau } \right)}^2}}}} \right) = 
\frac{{Dp}}{{{p^\infty }}},\\
\end{gathered}
\end{equation}
where for the sake of simplicity we set $\left\{ {\frac{r}{{{R_0}}},\frac{t}{{{t_0}}}} \right\} \equiv 
\left\{ {\chi ,\tau } \right\}$ and we achieved standard non-dimensionality of the partial differential 
equation by setting $\frac{\rho^\infty }{{{p^\infty }}}{\left( {\frac{{2{R_0}}}{{{t_0}}}} \right)^2} \equiv 
1$. With a similar procedure  the velocity field is:
\begin{equation}\label{Eq11}
\frac{{{v_r}}}{{{v_0}}} = \frac{{2\tau {{\left( {1 + {\tau ^2}} \right)}^2}}}{{{\chi ^2}}}.
\end{equation}
with ${v_0} \equiv \frac{{{R_0}}}{{{t_0}}}$.  In this way, the expansion of the convective elements increases 
with time  $\chi \propto {\tau ^2}$ while it is immediately evident that to the leading order (i.e., by series 
expansion to $\partial L$) the pressure \textit{difference}, $Dp$, drops with radius but \textit{increases } 
with time as $Dp \propto {\tau ^2}$ to follow the Bernoulli theorem. In particular the following 
\textit{temporal} relation holds good:
\begin{equation}\label{Eq12}
\chi  \propto {\tau ^2} \Rightarrow Dp \propto {\tau ^2},
\end{equation}
which correlates the temporal evolution of the  pressure at the convective element surface to that of the 
medium surrounding the convective element  as imposed and predicted by Eq.\eqref{Eq03}.
To better clarify the results we have obtained, we plot in Fig.\ref{FigPressure} and Fig. \ref{FigVelocity} 
the temporal and spatial dependence of the pressure and velocity fields around ideal convective elements as 
obtained in Eq.\eqref{Eq10} and \eqref{Eq11}.

In Fig.\ref{FigPressure} the different lines represent the different temporal evolution of the pressure 
field. 
As time increases, the convective element grows in size as given by Eq.\eqref{Eq09} and represented by the 3D 
spheres. It is soon evident that not only $Dp \ne 0\forall t$, but even grows with time, contradicting the 
widespread notion that $Dp = 0$ \citep[e.g., Eq.(6.2) in ][]{2012sse..book.....K}. This is, as noted in 
Sec.\ref{SecIntro}, indeed required from Eq.\eqref{Eq03}, the fundamental equation of convective turbulence. 
Only infinitely far from the convective element $Dp \to 0$, but this simplified scheme (and SFCT) fails 
because a stellar layer is not infinite nor it does last infinite time(\footnote{It also fails mathematically 
because $L$ has been assumed to be compact (finite in space and time).}). Since in this example, no 
hydrostatic equilibrium has been imposed,   ${p^\infty }$ can be the pressure existing at the surface of any 
nearby element. We also plotted for comparison in Fig.\ref{FigPressure} the spatial pressure dependence 
expected far away from the convective element in the case of rotational media \citep[dashed 
line,][]{2000ifd..book.....B}.

Similar considerations can be made for the velocity field surrounding a convective element as a function of 
time and position (see Fig.\ref{FigVelocity}). Note that the velocity of any fluid particle at the arbitrary 
location $\bm{r}$, ${v_r}={v_r (\bm{r};t)} \ne \dot R$ everywhere apart from the instantaneous location of the 
sphere overlapping the convective element surface as evident from Fig.\ref{FigVelocity}.

This case we have just illustrated is useful when compared to the more complex treatment presented in P14, 
and it helps to understand the meaning of the decoupled equations of the motion presented in Eq.(3) of P16 
mentioned in the previous section. A few points are worth highlighting:
\begin{enumerate}
	\item The pressure gradients presented in this simple case show the same asymptotic trends displayed 
in the more complex, non-inertial treatment of moving convective elements used by P14. At first glance, the 
ability of this simple exercise of capturing the essence of the SFCT  may be somewhat surprising not only for 
the different formalism but also the different physical content.  The SFCT \textit{requires} both the 
hypotheses of hydrostatic-equilibrium far away from the convective element and the condition $\left\| 
{{{{\bm{\dot R }}}}} \right\| \gg {v_{O'}}$, otherwise, the quadratic dependence of Eq.(\eqref{Eq12}) is not 
formally met. Here neither is used, but the plot of Fig. \ref{FigPressure} is very similar to that of Fig.1 in 
P16 where a similar trend was derived.
	
	\item  The velocity field  shown in Fig. \ref{FigVelocity} tells us that the longer the elapsed time, 
the more the convective element expands/contracts, the less the approximation $Dp = 0$ holds true. The 
relationship between the radial dimension and time for the generic convective element,  see  Eq.\eqref{Eq09}  
above, can be compared to which of Eq.(B15) of P16 that presents the same trend.
	
	\item The velocity-potential formalism yields the trend of pressure across the medium and does not 
affect the vertical motion. The decoupling of the independent coordinates demonstrated by P16 (their Eq.3) 
using arguments based on Classical Mechanics about translation and expansion of convective elements is here 
clarified from hydrodynamic arguments based on the Bernoulli equation (Eq.\eqref{Eq04}).  This result 
strengthens the Corollary 1 in Section 4.2 of P14: the hydrodynamics of a single thin stellar layer can be 
considered subject to a constant acceleration throughout it (as far as Eq.\eqref{Eq03} is concerned).
	
	\item The temporal evolution of the pressure and expansion of the convective elements are intimately 
related and cannot be separated without violating the energy conservation principle  (i.e., the Bernoulli 
theorem). If we want to examine the expansion/contraction of the convective elements, we cannot ignore the 
time evolution of the pressure and vice versa. This is done coherently in the SFCT but not in the MLT. The MLT 
copes with its fundamental failure in describing the temporal evolution of the system by introducing a free 
parameter that tacitly accounts for the conservation of the energy as a function of time.
	
	\item The hypothesis  $Dp = 0$ is \textit{not} at the basis of the Schwarzschild and Ledoux criteria 
for instability, ${\nabla _{{\text{rad}}}} < {\nabla _{{\text{ad}}}} + \frac{\varphi }{\delta }{\nabla _\mu }$ 
as already proved in Lemma 2 of P14 and where  $\nabla _{{\text{rad}}}$, $\nabla _{{\text{ad}}}$, 
${\nabla _\mu }$ are the classical radiative, adiabatic and molecular weight gradients as defined in 
Sec.\ref{SecEssence}.
	
\end{enumerate}

To summarize, what emerges from Figs.\ref{FigPressure} and \ref{FigVelocity} is that as the bubble expands, 
the pressure difference with the surrounding medium moves further away from equilibrium, rather than towards 
it.
In this illustrating case where there is no translational movement of the bubble, the increase in size which 
is driving this evolution is put in by hand through Eq.\eqref{Eq09}, so that there is no need for concern at 
the source of energy to drive this process. In the complete treatment of P14 which encompasses the movement of 
the bubble, this expansion is self-consistently derived, and it is efficiently driven by the gradient of 
temperature and pressure gradients $\nabla$ and $\nabla_e$ when a layer $L$ becomes convective. We provide an 
illustrating check on this process in Sec.\ref{example}. While the eddy expansion is taking place, 
instabilities set in, destroying the bubble, as discussed in Sec.\ref{SecFate} below.

\subsection{The danger of the instantaneous pressure readjustment}
Failing to understand Eq.\eqref{Eq03}, whose original analysis dates back to the milestone works of 
\citet[][]{ 1956RSPTA.248..369B, 1967JFM....27..581S},  as well as failing the energy conservation required by 
the Bernoulli equation, can lead to incorrect results as in the recent work of \citet[][from now on 
MB16]{2016MNRAS.457.4441M}. While in the MLT the fundamental relation in Eq.\eqref{Eq03} is implicitly hidden 
by a free parameter ${\Lambda _m}$, it is explicitly not fulfilled  in the recent study of MB16. In light 
of the work reviewed in Sec.\ref{SecIntro} and the example in Sec.\ref{SecHomoSFCT}, the analysis of the 
potential-vector approximation presented by MB16 is incorrect on three counts:
\begin{itemize}
	\item the authors assume instantaneous hydrostatic equilibrium on the single bubble treatment. 
This is evident in the derivation of their Eq.(2)  which does not contain  any new free parameter as is 
provided by the MLT to compensate for the violation of the basic relationship of Eq.\eqref{Eq03}. In contrast, 
MB16 writes $$\frac{{dp\left( {{\bm{ r }};t} \right)}}{{dt}} = \frac{{dp\left( {\bm{ r }} 
\right)}}{{dr}}{v_b}\left( t \right)$$
so that for ${v_b}\left( t \right) = 0$, i.e. a  bubble at rest, one has
$$\frac{{dp\left( {{\bm{ r }};t} \right)}}{{dt}} = 0 \Leftrightarrow p\left( {{\bm{ r }};t} \right) = p\left( 
{\infty ;t} \right) = \text{cnst.}$$
However, this contradicts the fundamental equation of convection Eq.\eqref{Eq03} that shows how the 
pressure at the surface of the bubble and infinity is never the same. The MB16 approach leads to their Eq.(8) 
that to our knowledge has no regime of validity. Later with their Eq.(A6) MB16 claim to prove that a 
convective element and medium have almost the same pressure. As well known since the early studies on the 
pressure fields around an eddy by \citet{1956RSPTA.248..369B} and \citet{1967JFM....27..581S} \textit{this is 
never the case}. The works of \citet{1956RSPTA.248..369B} and \citet{1967JFM....27..581S} are well known and 
influential results that had a wide range of implications in physics and astrophysics 
\citep[e.g.,][]{1966JFM....25...97R, 
1967PhFl...10.1224L,1977PhRvA..16..732F,1981JFM...106...27M,1983PhFl...26.2080B,1990JFM...212..497H,
1991PhRvL..67.3243K,1992PhFlA...4.1492G,1996JFM...318..303H,2000PhFl...12.1997S,2000JTurb...1...10O,
2002AnRFM..34...19M,2010PhRvE..81a6316P, 2010JFM...663..268D, 2011JFM...668..351M, 2012JFM...706..150D, 
2012PhRvE..86f6320T, 2014JFM...738..378K, 2014PhRvE..90d3019Y, 2015PhFl...27i5103S, 
2017A&A...599A..69R,2017IJNLM..95..143L,2017PhR...720....1Z}, over which SFCT stands, and that here we just 
proposed again in the example of Fig.\ref{FigPressure}, where convective elements and medium have 
approximately the same pressure only for, say, $\tau \leq 1$ and never more as time passes(\footnote{Note that 
Fig.\ref{FigPressure} refers to the exercise of Sec.\ref{SecEssence}. To obtain the correct numerical values 
for densities, temperatures, times, etc., the layer under examination has to be inserted into an environment 
given by an EoS of the star. This was already shown in Fig.1 of P16 and not repeated here.}).
	\item By construction, no equilibrium can exist for a convective element coming into existence. 
The authors assume instantaneous adjustment of the  pressure equilibrium  (see their Eq.(8)), i.e., MB16 do 
not take into account time/space evolution of pressure, but at the same time they follow the temporal 
evolution of the bubble size $\left\| {{{\bm{x}}_e}\left( t \right)} \right\|$ (one of the degrees of freedom 
of the system in ${S_O}$) and the motion ${{\bm{x}}_O}\left( t \right)$ (the other degree of freedom), which 
is mathematically and physically inconsistent.
	\item No turbulence can exist if the pressure adjusts itself instantaneously: the pressure waves 
caused by the pressure fluctuations at a given eddy position are indeed the trigger for the formation of other 
eddies and give rise to turbulent convection. While SFCT captures the fundamentals of this process, it is 
missing in MB16 (see, e.g., their result in Appendix Eq.(A6)).
\end{itemize}

The most obvious problem in MB16 is clear after our Eq.\eqref{Eq12} (repeated here for simplicity):
$$
\chi  \propto {\tau ^2} \Rightarrow Dp \propto {\tau ^2},
$$
at the surface of the convective element. Contrary to what shown here in compliance with the physics 
expressed 
by Eq.\eqref{Eq03},  MB16 derive their Eq. (A6) (which  sustains their Eq.(2) and the rest of the paper)  that 
predicts the opposite of our  Eq.\eqref{Eq12}. The authors fail to investigate the limit of their Eq.(A6) 
because they compute the limit of $p = p\left( {r,t} \right)$ without examining whether the bubble exists in 
that regime in time and space.
We have shown here that if Bernoulli's theorem is correctly fulfilled (and hence the energy conservation that 
it represents), it is never the case that ${p^\infty } \simeq p$ anywhere and at any time.
Coherently, the velocity field is shown in Fig. \ref{FigVelocity} tells us that: the longer the elapsed time, 
  the more the convective element expands/contracts, the less the approximation of Eq. (A6) in MB16 is 
verified.
Indeed the correct physical formulation and analysis of the whole problem impose the pressure-time dependence 
of Eq.\eqref{Eq12}. In particular even far away from the bubble the condition ${p^\infty } \simeq p$  cannot 
happen because of the temporal divergence in the equations (in any case convective bubbles are destroyed by 
instabilities as examined in \citealt[][]{2015A&A...573A..48P} as time goes by)(\footnote{It can be proved by 
investigation of  $\chi  = 1 + {\tau ^n}$ (with $n \in \mathbb{Z}$ ) that only for $n = 1$ the pressure at the 
surface $r = R\left( t \right)$ remains approximatively constant (it results in that $Dp \in \left[ {1.4,0.0} 
\right[$). This seems to be the only mathematical case in which the study of MB16 may be applicable or in 
which the immediate pressure balance between convective elements and the medium is established  \citep[see for 
instance ][ their  Eq. 6.2]{2012sse..book.....K}  even though, to our knowledge,  no evidence exists where the 
proportionality $\chi  = 1 + \tau$ is met.
}).

\subsection{The fate of  a convective element}\label{SecFate}
Finally, we examine the fate of an expanding/contracting convective element.

As already mentioned in P16 the SFCT fails in compressible regimes and in the very last layer of the star 
where 
suitable boundary conditions must be supplied(\footnote{Note how in Fig.\ref{FigSchemeTime} caption we exclude 
the layer $L_n$ for $n=M$.}). Also, convective elements cease to exist at $t \to \infty$. The reason for this 
failure of the theory at significant times and very outer regions is that in our analysis we did not include 
the instabilities that dissipate and destroy the elements.
Possible  physical causes of disintegration include:

\begin{enumerate}
	\item Deformation: the convective elements dissolve, losing their initial spherical geometry. This 
reflects the classical picture of a turbulent eddy that winds up in vortex sheets in a sequence of azimuthal 
vorticity and poloidal motions and sweeps angular momentum outward radially to form sheets. The start of this 
process is captured by the linear treatment regarding the stability parameter $\gamma _I^2 = \gamma _I^2\left( 
{{\bm{x}};t} \right)$  already available in classical literature on the subject 
\citep[e.g.,][]{1954JAP....25...96P}.
	
	\item Rayleigh-Taylor instability: finger-like penetrations of two fluids at different densities 
(e.g., a convective element expanding inside a radiative/convective fluid). The process depends on (i) the 
density difference between the two fluids; the instability parameter is $\gamma _{RT}^2 = \gamma _{RT}^2\left( 
{{\bm{x}};t} \right)$, and  (ii) the relative acceleration between the two fluids with instability parameter  
$\gamma _{a - RT}^2 = \gamma _{a - RT}^2\left( {{\bm{x}};t} \right)$. See \ref{AppendixA} for more 
details.
	
	\item Kelvin-Helmholtz instability: the relative sliding of the inner layers concerning external ones 
of convective the elements helps to dissolve the convective element themselves. The instability parameter is   
$\gamma _{KH}^2 = \gamma _{KH}^2\left( {{\bm{x}};t} \right)$.  To the first order,  no dependence on the 
acceleration of the convective element is found. See \citealt[][]{2015A&A...573A..48P}  for an extensive 
investigation and Appendix A for more details(\footnote{Note that the theory is presented in  P16 for the 
specific case of a cloud at much higher temperature moving through an ionized plasma. However, the theory 
applies to any non-degenerate, non-relativistic gas in a convective zone inside a star.}).
\end{enumerate}
A convenient analytical solution is possible using the WKB approximation and a simple formulation for the 
linear-instability parameter results as:
\begin{equation}
{\gamma ^2} = \gamma _I^2 + \gamma _{RT}^2 + \gamma _{a - RT}^2 + \gamma _{KH}^2 + \gamma _{\rm{mix}}^2,
\end{equation}
see Eq.\eqref{EqA5}. Here $\gamma _{\rm{mix}}^2$ refers to the linear superposition of the instability effects.

\section{Numerical results of SFCT}\label{example}
After getting a better insight into the SFCT paradigm by studying how its behaves in the homogeneous 
isotropic 
case and in respect to the MLT, we pass now from the single eddy description to the collective averaged 
behavior of a system of convective cells embedded in a real stellar layer.

The generic layer $L_n$ is always assumed to be in hydrostatic equilibrium both in the MLT and in the SFCT
even though $L_n$ is a sum of elements not in hydrostatic equilibrium (i.e., the convective elements). This 
description differs from the test-cases of Sec.\ref{SecEssence} because the single equation describing one 
eddy element is now embedded in a system of equations that describe the layer $L_n$  and the interaction of 
all the eddies with this layer.
This single-to-collective description is achieved solely through a parameter (the mixing length ${{\Lambda 
}_{m}}$) 
in the MLT. The resulting set of equations is reported here (see, e.g., Eq.(B1) of P16):
\begin{equation}\label{MLTsys}
\left\{ \begin{aligned}
{{\varphi }_{\text{rad }\!\!|\!\!\text{ cnd}}} & = \frac{4ac}{3}\frac{{{T}^{4}}}{\kappa {{h}_{p}}\rho }\nabla   \\
{{\varphi }_{\text{rad }\!\!|\!\!\text{ cnd}}}+{{\varphi }_{\text{cnv}}} & = \frac{4ac}{3}
\frac{{{T}^{4}}}{\kappa {{h}_{p}}\rho }{{\nabla }_{\text{rad}}}  \\
{{v}^{2}} & = g\delta \left( \nabla -{{\nabla }_{e}} \right)\frac{l_{m}^{2}}{8{{h}_{p}}}  \\
{{\varphi }_{\text{cnv}}} & = \rho {{c}_{P}}T\sqrt{g\delta }\frac{l _{m}^{2}}{4\sqrt{2}}h_{p}^{-3/2}
{{\left( \nabla -{{\nabla }_{e}} \right)}^{3/2}}  \\
\frac{{{\nabla }_{e}}-{{\nabla }_{\text{ad}}}}{\nabla -{{\nabla }_{e}}} & = \frac{6ac{{T}^{3}}}
{\kappa {{\rho }^{2}}{{c}_{p}}{{l}_{m}}v},  \\
\end{aligned} \right.
\end{equation}
The same connection single-to-global description can also be achieved in the SFCT with this set of equations 
(see, e.g., Eq.(B21) of P16):
\begin{equation}\label{SFCTsys}
\left\{ \begin{aligned}
{{\varphi }_{\text{rad/cnd}}} & =  \frac{4ac}{3}\frac{{{T}^{4}}}{\kappa {{h}_{p}}\rho }\nabla   \\
{{\varphi }_{\text{rad/cnd}}}+{{\varphi }_{\text{cnv}}} & =  \frac{4ac}{3}\frac{{{T}^{4}}}
{\kappa {{h}_{p}}\rho }{{\nabla }_{\text{rad}}}  \\
\frac{{{{\bar{v}}}^{2}}}{R} & =  4{{g}_{4}}\frac{\nabla -{{\nabla }_{e}}-\frac{\varphi }{\delta }
{{\nabla }_{\mu }}}{\frac{3{{h}_{p}}}{2\delta \bar{v}\tau }+{{\nabla }_{e}}+2\nabla -\frac{\varphi }{2\delta 
}{{\nabla }_{\mu }}}  \\
{{\varphi }_{\text{cnv}}} & =  \frac{1}{2}\rho {{c}_{p}}T\left( \nabla -{{\nabla }_{e}} \right)
\frac{{{{\bar{v}}}^{2}}\tau }{{{h}_{p}}}  \\
\frac{{{\nabla }_{e}}-{{\nabla }_{\text{ad}}}}{\nabla -{{\nabla }_{e}}} & =  
\frac{4ac{{T}^{3}}}{\kappa {{\rho }^{2}}{{c}_{p}}}\frac{\tau }{{{R}^{2}}}  \\
R & =  {{g}_{4}}\bar{\chi }\frac{\nabla -{{\nabla }_{e}}-\frac{\varphi }{\delta }{{\nabla }_{\mu }}}
{\frac{3{{h}_{p}}}{2\delta \bar{v}\tau }+{{\nabla }_{e}}+2\nabla -\frac{\varphi }{2\delta }{{\nabla }_{\mu 
}}}.  \\
\end{aligned} \right.
\end{equation}
Here all the quantities are averaged on their values at the generic layer ${{L}_{n}}\left( \mathbf{x} 
\right)$ 
of Fig.\ref{FigSchemeTime}. In particular, for the sake of notation, we simplify the notation for $T\left( L 
\right)=T\left( \mathbf{x} \right)={{T}^{\infty }}$ in  $T$ for the temperature, $p^{\infty }$ in $p$ for the 
pressure, and so forth, for density $\rho $, opacity $\kappa $, specific heat at constant pressure 
${{c}_{p}}$, pressure scale length ${{h}_{p}}$, the gravity ${{g}_{4}}=g/4$, radiative/conductive flux 
${{\varphi }_{\text{rad/cnd}}}$, convective flux $\varphi_{\text{cnv}}$, adiabatic gradient 
$\nabla_\text{ad}$, radiative gradient $\nabla_{\text{rad}}$; and where we adopt the standard notation $\delta 
\equiv -{{\left. \frac{\partial \ln \rho }{\partial \ln T} \right|}_{p,\mu }}$, $\varphi \equiv {{\left. 
\frac{\partial \ln \rho }{\partial \ln \mu } \right|}_{p,T}}$  with $\mu $ molecular weight. Here, $a$ is the 
radiation pressure constant and $c$ the speed of light.

Note how the mathematical nature of these two systems is profoundly different. In the MLT case, the system
 of Eq.\eqref{MLTsys} is local and time independent (i.e. pertinent the layer $L_n$ at every time): it can be 
solved at every location $\mathbf{x}$ where the defining quantities of $L$ result well defined. The case of 
the SFCT system of equation is intrinsically more complex because of its time dependence. In 
Eq.\eqref{SFCTsys} enter both time and location dependent quantities, i.e. the equations of motion for the 
average eddy, together with time-independent but location-dependent quantities, i.e. the equations related to 
the  Schwarzschild or Ledoux criteria and to the radiation transfer processes(\footnote{This sets the SFCT 
in that branch of Mathematics called Differential Algebra \citep[e.g.,] [see also P14]{2015DAS} and its 
defining system is indeed an algebraic-differential system  (DAS) of equations. The reduction of the original 
DAS to the system of Eq.\eqref{SFCTsys} is achieved in P14. In the case of the Eq.\eqref{SFCTsys} a theorem 
(Theorem of uniqueness, P14) grants that the solution of the system in terms of the mean stream velocity 
$\bar{v}$, convective flux ${{\varphi }_{\text{cnv}}}$, radiative/conductive flux ${{\varphi 
}_{\text{rad/cnd}}}$, stellar gradient $\nabla $, and eddy gradient ${{\nabla }_{e}}$, exists and it is unique 
once the layer $L=L\left( T,\kappa ,\rho ,{{\nabla }_{\text{rad}}},{{\nabla }_{\text{ad}}},{{\nabla }_{\mu 
}},g,{{c}_{p}} \right)$ is assigned (with $\bar{\chi }$ monotonic function of the parameter $\tau $). The 
simple algebraic system of equation for the MLT is studied elsewhere \citep[e.g.,][]{1994sse..book.....K}.}).

\subsection{A simple numerical validation of the SFCT}
We present here a simple numerical test of whether the SFCT can eliminate the mixing-length and why the 
uniqueness theorem presented in P14 holds good. As a result of this test, not only may we better understand 
the theory of  P14 but also a numerical counterpart will be available to the analytical derivation of the set 
of equations presented by P16. The test does not require complicated stellar structure codes; instead, it can 
be followed with a pocket calculator.

We choose the same layer of the standard solar model already examined by P14. The layer is close to the 
surface
 where energy transport is far from being super-adiabatic, hence where the effect of the SFCT (or MLT) is 
more substantial and relevant. We use the same input values for our calculation in both MLT and SFCT and adopt 
SI units. The input values are,
$r = 6.92002 \times {10^8}\, \text{m}$,
$T = 48503.4\, \text{K}$,
${\nabla _{{\text{ad}}}} = 0.28310$,
${\nabla _{{\text{rad}}}} = 0.295186$,
${p^\infty } = 2.05058\, \text{N} \, {\text{m}^{ - 2}}$,
$g = 277.517\, \text{m} \, {\text{s}^{ - 2}}$,
$\rho  = 0.345519 \, \text{kg} \,  {\text{m}^{-3}}$,        
$c_p^\infty  = 102986.2 \,\text{J} \,  \text{kg}^{ - 1} \, \text{K}$
for stellar radius, temperature, adiabatic gradient, radiative gradient, pressure of the stellar layer far 
away from the convective element, gravity in the same layer, average density and specific heat at constant 
pressure far away from the convective element respectively. In addition  to these values, three universal 
constants must be specified:
gravity constant
$G = 6.67428 \times {10^{ - 11}}\,{\text{m}^3} \, \text{Kg}^{ - 1} \, {\text{s}^{ - 2}}$,
the gas radiation density constant
$a = 7.56464 \times {10^{ - 16}}\, \text{J} \, {\text{K}^{ - 4}} \, {\text{m}^{ - 3}}$
and the speed of light
$c = 2.99792 \times {10^8}\, \text{m} \, {\text{s}^{ - 1}}$.

We first obtain a solution for the radiative-convective-conductive transport of energy with the MLT. For this 
we use Eqs.(B1), (B2) and (B3) of P16. To proceed we need  the mixing-length of ${l_m} = {\Lambda _m}{h_p}$  
with ${\Lambda _m} = 1.64$ and ${h_p} = \frac{{{p}}}{{\rho g}}$  \citep[e.g.,][]{2008A&A...484..815B}. With 
these entries, we can define the quantities $V$ and $W$ of  Eq.(P16,B2), that we report here for simplicity 
and we evaluate as:
$$V = \frac{{3ac{T^3}}}{{c_p {\rho ^2}\kappa l_m^2}}\sqrt {\frac{{8{h_p}}}{{g\delta }}}  = 1.95043
 \times {10^{ - 11}}$$
and
$$W = {\nabla _{{\text{rad}}}} - {\nabla _{{\text{ad}}}} = 295186.3. $$
The expression for Eq.(P16,B3) becomes
\[\begin{gathered}
{\left( {x - 1.950 \times {{10}^{ - 11}}} \right)^3} +  \hfill \\
+ \left( {1.733 \times {{10}^{ - 11}}} \right)\left( {{x^3} - 2.951 \times {{10}^5}} \right) = 0 \hfill \\
\end{gathered} \]
with
$${x^2} = \nabla  - {\nabla _{{\text{ad}}}} + {V^2}.$$
This is the classic cubic equation of the MLT. Using now Eq.(B4), (B5) and (B6) of P16 one can derive the  
analytical solution. Alternatively, a numerical solution can be used \citep[e.g.,][]{1992nrfa.book.....P} to 
obtain $x = 0.01723$ for our layer. Finally, with this value for $x$ we derive the layer pressure gradient as 
$\nabla  = 0.28340$.

We repeat now the same calculations but using the SFCT for which no assumption has to be made for the mixing 
length parameter  ${\Lambda _m}$ simply because this parameter is missing in the equations of this theory. We 
start from Eq.(12) of P16, and we proceed backward to obtain  $\nabla$. We use the same input values and take 
time interval sufficiently long, say $t>{{t}^{(\infty) }}$, to ensure that the asymptotic regime is reached 
for all relevant quantities (the velocity in particular)(\footnote{It is worth recalling here that the 
asymptotic regime is a consequence of the uniqueness theorem (see P14, Sec.6.2). By itself, the single eddy 
equation of motion does not predict any asymptotic value. In particular, the SFCT does not apply at all in the 
case of a bubble rising in a medium and reaching the terminal velocity of traditional mechanic systems. As 
shown in Sec.\ref{SecFate}, the convective cell dissipate far before reaching any terminal velocity by 
instability effects. The asymptotic regime evidence in P14 and P16 (and hereafter referred to with 
$t>t^{(\infty)}$) manifests itself only when we embed the averaged single equation of motion for the 
convective eddy inside the system of equation of the convective layer. The solution of those six equations in 
Eq.\eqref{SFCTsys} (i.e., once simultaneously considered) presents an asymptotic behavior; by itself the 
single equation of motion of SFCT does not.}).
From a numerical point of view, we consider the asymptotic regime to be reached when the velocity of 
convective 
elements no longer change more than a few percent (the corresponding integration time is about $t < \infty 
\equiv t^{(\infty)} = {10^6}$ s).   The basic quintic equation of the system Eq.\eqref{SFCTsys} is given 
by(\footnote{A note on the notation: here all the quantities of the SFCT (as well as for the MLT) refer to the 
average behavior over the layer $L$. Nevertheless, only for the velocities $\bar v$ and the expansion function 
$\bar \chi$ we use an upper bar to distinguish them from the single bubble value in order not to confuse them 
with the previous section notation. The process of average for $\bar v$ and $\bar \chi$ is detailed in P14 and 
in the next section too.})
\begin{equation}\label{Quintic}
\sum\limits_{i = 0}^5 {{c_i}{\bar v^i} = 0},
\end{equation}
whose  coefficients are, and take the numerical values:
\begin{eqnarray}\label{QuinticCoeff}
{c_5} &=& 1, \nonumber \\
{c_4} &=& \frac{{{h_p}}}{{2\delta {\nabla _{{\text{ad}}}}\tau}} = 37.7687,\nonumber \\
{c_3} &=& \frac{{8\alpha ({\nabla _{{\text{ad}}}} + 2{\nabla _{{\text{rad}}}})}}{{9{\nabla 
_{{\text{ad}}}}\tau}} = 3.75137, \nonumber \\
{c_2} &=& \frac{{4\alpha \left( {2\delta {g_4}\tau({\nabla _{{\text{ad}}}} - 
{\nabla _{{\text{rad}}}})\sqrt {\bar \chi}   + 3{h_p}} \right)}}{{9\delta {\nabla _{{\text{ad}}}}{\tau ^2}}}  
=  3.75 \times 10^6, \nonumber \\
{c_1} &=& \frac{{32{\alpha ^2}{\nabla _{{\text{rad}}}}}}{{3{\nabla _{{\text{ad}}}}\bar\chi }} = 0.000546,  
\nonumber \\
{c_0} &=&  \frac{16 \alpha ^2 h_p}{3 \delta  \nabla _{\text{ad}} \tau  \bar\chi } = 1.97995 \times 10^{-8},
\end{eqnarray}
with
\begin{equation}\label{Alpha}
\alpha\equiv\frac{a c T^3}{\kappa  \rho ^2 c_p}.
\end{equation}
This equation has three real solutions and two complex ones. The only physically acceptable solution is 
$\bar v = 194.607 \,\text{m} \, {\text{s}^{ - 1}}$. After the selection of the roots, the determination of 
$\nabla$ becomes straightforward.
Finally, we obtain ${\nabla _{{\text{SFCT}}}} = 0.28310$ for the stellar layer under examination which is 
virtually indistinguishable from the result above of the MLT: ${\nabla _{{\text{MLT}}}} = 0.28340$.
The ``mixing-length problem'' is numerically solved: equalities between the left-hand side (LHS) and 
right-hand 
side (RHS) of the system of equations are recovered and verified (within the precision error) and the Sun 
modelled with MLT and with SFCT is the same, i.e., it has the same star pressure-temperature stratification.

\subsection{ Time dependence Schwarzschild or Ledoux criteria}
\begin{figure*}
	\includegraphics[width=180mm]{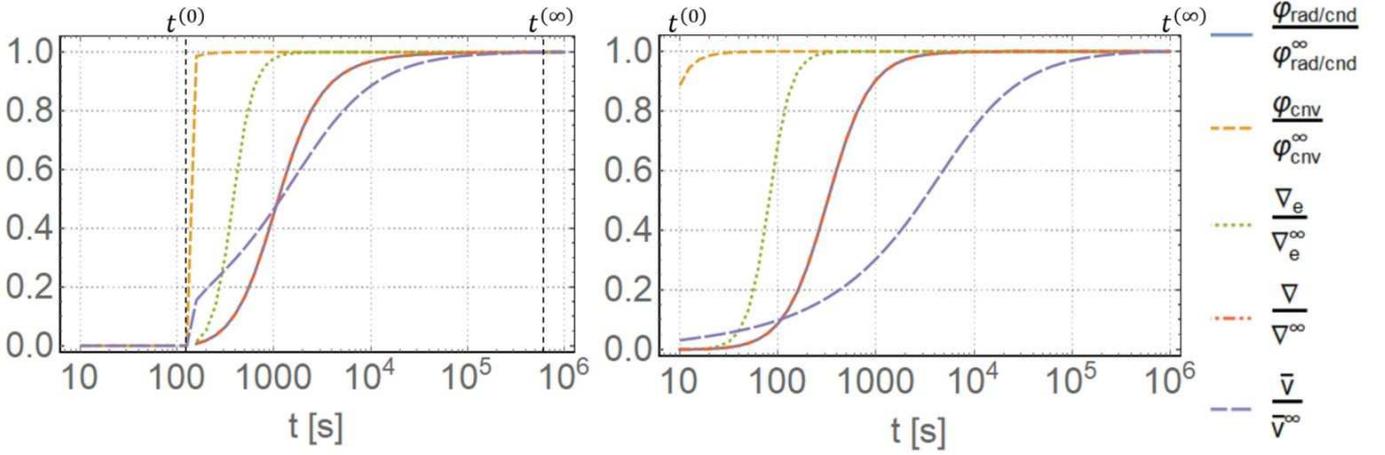}   
	\caption{Evolution of the averaged fluxes, gradients and velocity for a $1 M_\odot$ stellar model 
with ${\log _{10}}L/{L_ \odot } = 0.0$, ${\log _{10}}{T_{{\rm{eff}}}} = 3.76$ and solar metallicity. Left 
panel refers to $R = 0.7 R_{\text{max}}$, right panel to $R = 0.88 R_{\text{max}}$. We refer to $t<t^{0}$ as 
the homogeneous isotropic turbulence regime treated in Sec.\ref{SecEssence}, to $t^{(0)}<t<t^{(\infty)}$ as 
the transition regime, and to $t<t^{(\infty)}$ as the asymptotic regime.}
	\label{FigTimescale01}
\end{figure*}

\begin{figure*}
	\includegraphics[width=180mm]{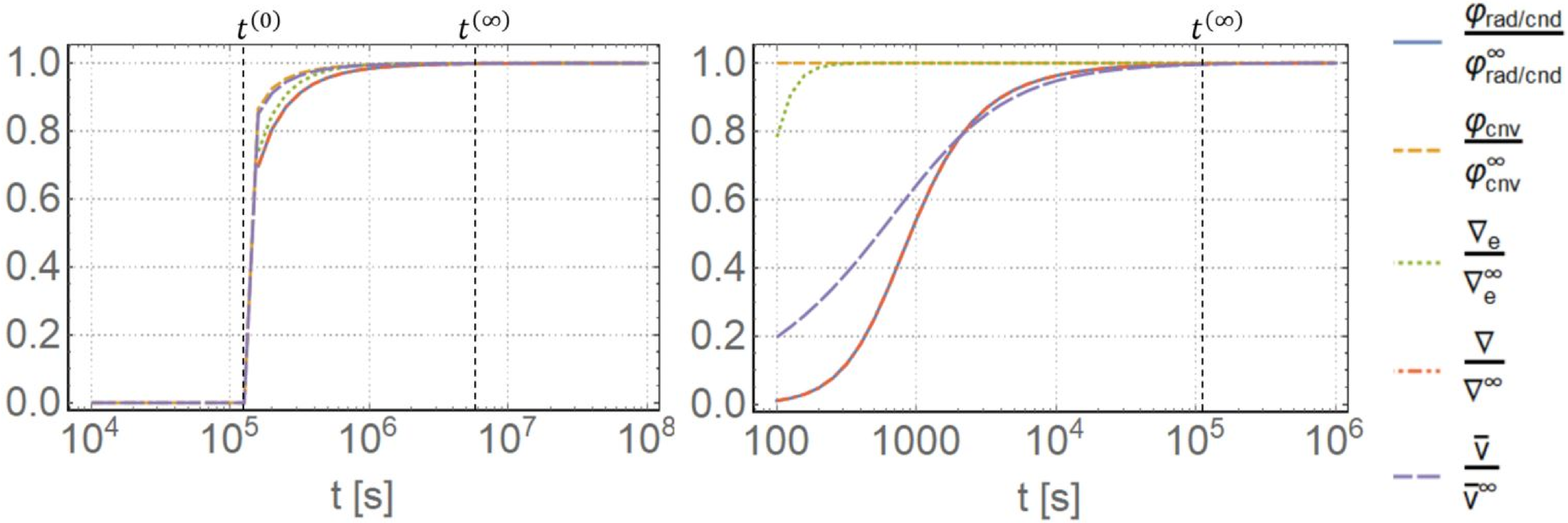}   
	\caption{Same as Fig.\ref{FigTimescale01} but for stellar model 
with ${\log _{10}}L/{L_ \odot } = 0.3$, ${\log _{10}}{T_{{\rm{eff}}}} = 3.70$ }
	\label{FigTimescale02}
\end{figure*}

\begin{figure*}
	\includegraphics[width=180mm]{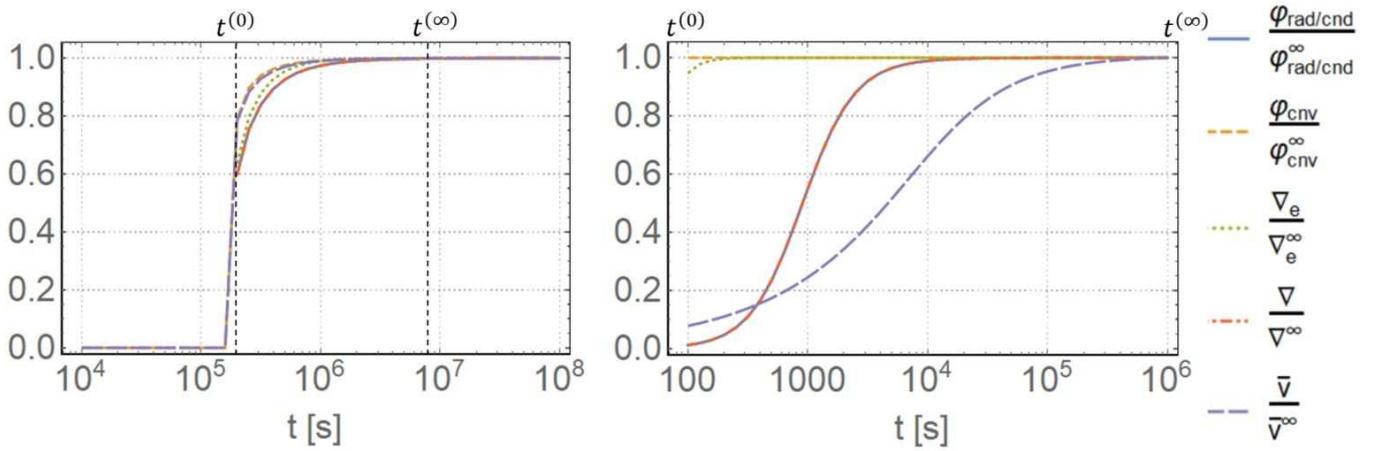}   
	\caption{Same as Fig.\ref{FigTimescale02} but for stellar model 
with ${\log _{10}}L/{L_ \odot } = 2.0$, ${\log _{10}}{T_{{\rm{eff}}}} = 3.60$}
	\label{FigTimescale03}
\end{figure*}

By extending the detailed study of the stellar model to different phases of a solar track as in P16, we 
obtain 
the Figs. \ref{FigTimescale01}, \ref{FigTimescale02}, and \ref{FigTimescale03} from which to extract a few 
essential features of the SFCT treatment.

While, generally, the stability criteria of Schwarzschild or Ledoux express the condition for the onset of
 the convection, they do not predict the timescale for a layer becoming convective. This is because they are 
not time-dependent criteria. The SFCT being a time-dependent theory, theory predicts a non-zero time-lapse 
for 
the convection to become effective. For the sake of simplicity, we refer to a stellar layer $L$  as 
"convective," if (and only if) both the conditions
\begin{equation}\label{Criteria}
\left\{ \begin{gathered}
{\nabla _{{\text{rad}}}} > \nabla  > {\nabla _e} > {\nabla _{{\text{ad}}}} \hfill \\
{\varphi _{\text{rad/cnd}}} < {\varphi _{{\text{cnv}}}}, \hfill \\
\end{gathered}  \right.
\end{equation}
are satisfied $\forall t$. Despite all the figures refer to layers where classic Schwarzschild criterion is
 satisfied, ${{\nabla }_{\text{rad}}}>\nabla >{{\nabla }_{e}}>{{\nabla }_{\text{ad}}}$, the onset of the 
convection is initially still radiative dominated energy transfer, i.e., ${{\varphi 
}_{\text{rad/cnd}}}>{{\varphi }_{\text{cnv}}}$, and only after a small time of the order of 100 seconds for a 
solar-like star (but which can be as large as ${{10}^{7}}$ s for a giant star) the energy flux becomes 
dominated by convection (i.e., both the criteria of Eq.\eqref{Criteria} are matched).
Finally, we note how the timescale with which 
$\nabla =\nabla \left( t \right)$ and ${{\varphi }_{\text{rad/cnd}}}={{\varphi }_{\text{rad/cnd}}}\left( t 
\right)$ develop is always the same. This is natural as they are directly proportional, ${{\varphi 
}_{\text{rad/cnd}}}=\frac{4c}{3}\frac{a{{T}^{4}}}{\kappa {{h}_{p}}\rho }\nabla$. Moreover, because of 
Eq.\eqref{Criteria}, we see that ${{\varphi }_{\text{rad/cnd}}}<{{\varphi }_{\text{cnd}}}$ in agreement with 
our definition of a convective layer.

Finally, the average velocity of the convective elements remains the more sensitive variable to the physics 
of the star. In our case, we are limited in Fig.\ref{FigTimescale01}, \ref{FigTimescale02}, and 
\ref{FigTimescale03} to  positive average velocities $\bar v>0$, but the analogous analysis can be done for 
negative velocities $\bar v<0$ with similar results. As evident from the figures, the timescale with which the 
asymptotic mean-stream velocity is reached by the convective elements, before dissolving by RT and KH effects, 
can be as short as a few hundred seconds as well as long as ${{10}^{7}}$ seconds for bigger stars.


\subsection{A numerical check}
Extending the numerical investigation introduced above, allows us to propose a robust numerical study of 
the subsonic regime approximations used in SFCT(\footnote{This corresponds indeed to a numerical validation of 
Lemma 1 in P14}). In particular, one of the standing hypothesis of SFCT, the limitation to subsonic regimes, 
translate to the condition
\begin{equation}\label{VF1}
{\left( {\frac{\bar v}{{{{\dot R }}}}} \right)^2}\frac{1}{2}\left( {\frac{9}{4}{{\sin }^2}\theta  - 1} 
\right) 
\ll A\frac{{{R }}}{{\dot R ^2}}\left( {\frac{3}{2}\cos \theta  - \cos \phi } \right) + \frac{{{{\ddot R }}{R 
}}}{{\dot R ^2}},
\end{equation}
or
\begin{equation}\label{VF2}
{\left( {\frac{{\bar v{{\dot R }}}}{{\dot R ^2}}} \right)^2}\frac{5}{2}\cos \theta  \ll A\frac{{{R }}}
{{\dot R ^2}}\left( {\frac{3}{2}\cos \theta  - \cos \phi } \right) + \frac{{{{\ddot R }}{R }}}{{\dot R ^2}},
\end{equation}
that we are able now to numerically verify over an extended set of models. We proceed as follows. For the 
sake of simplicity we define three functions from the above Eqs.\eqref{VF1} and \eqref{VF2}:
\begin{equation}
\begin{gathered}
f\left( {t,\theta } \right) \equiv {\left( {\frac{\bar v}{{{{\dot R }}}}} \right)^2}\frac{1}{2}
\left( {\frac{9}{4}{{\sin }^2}\theta  - 1} \right), \hfill \\
g\left( {t,\theta } \right) \equiv {\left( {\frac{{\bar v{{\dot R }}}}{{\dot R ^2}}} \right)^2}
\frac{5}{2}\cos \theta,  \hfill \\
k\left( {t,\theta } \right) \equiv A\frac{{{R }}}{{\dot R ^2}}
\left( {\frac{3}{2}\cos \theta  - \cos \phi } \right) + \frac{{{{\ddot R }}{R }}}{{\dot R ^2}}. \hfill \\
\end{gathered}
\end{equation}
We proceed to consider the same layer $L$ introduced in Sec.\ref{example}. Further values necessary for 
the computations of this exercise are available in Table 1 of P14 and the value of the derivative of $\bar v$, 
i.e. ${{{\dot R }}}$, are simply the tangents to the plot in Fig. \ref{FigTimescale01}, \ref{FigTimescale02}, 
or \ref{FigTimescale03}. We obtain for the same point inside the Sun analyzed in Sec.6.3 of P14 (after 
$\sim10^6$ s) that $\bar v = 194.6\;\text{km}\;{\text{s}^{ - 1}}$, $\bar \chi  = 8.3 \times {10^{10}}$ so that 
at $\theta  = 0$ gives $f\left( {{{10}^6},0} \right) =   - 6.81 \times {10^{ - 7}}$. At the same time and 
location we have $\dot{\bar{v}}=1.027\times {{10}^{-5}}\ \text{km}\ {{\text{s}}^{-2}}$, $\bar \chi ' = 1.666 
\times {10^5}$ and $\bar \chi '' = 0.156$ and hence $k\left( {{{10}^6},0} \right) = 0.5$. It is evident that 
the first of the equations necessary to prove the Lemma 1, $f\left( {t,\theta } \right) \ll k\left( {t,\theta 
} \right)$, of P14 is largely verified. Now that all the values are available, it is a simple exercise to 
prove that also the second $g\left( {t,\theta } \right) \ll k\left( {t,\theta } \right)$.

\begin{figure}
	\includegraphics[width=\columnwidth]{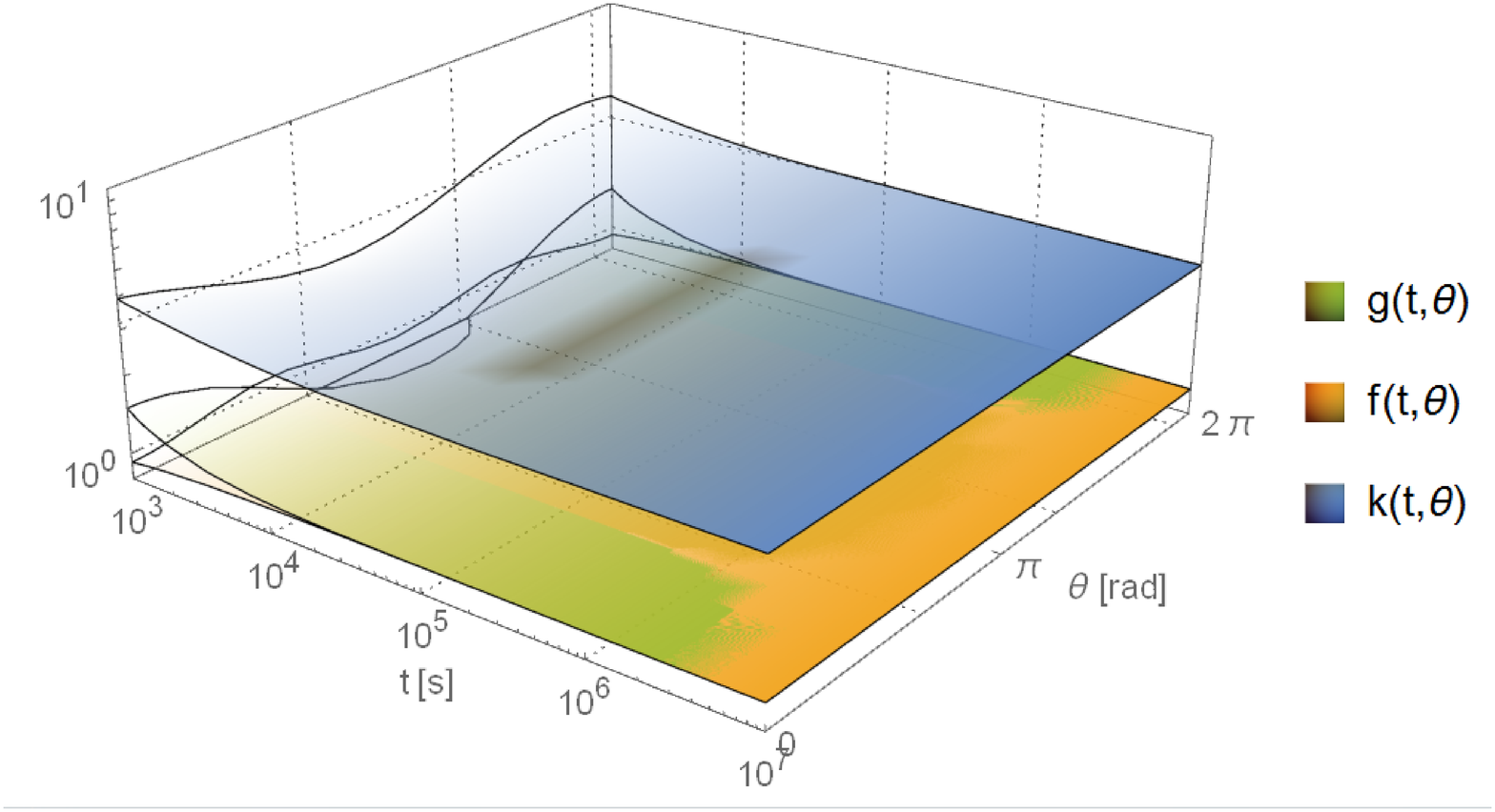}   
	\caption{Numerical validation of the proof of Eq.\eqref{VF1} and Eq.\eqref{VF2}. Functions are 
defined 
along the text and plotted against the values of interest, i.e. in the ``asymptotic'' regime of validity of 
the uniqueness theorem. As evident the blue manifold always overlook the green and yellow ones for all the 
time-scale of interest in the development of the convection in convective unstable layers and for all the 
angular dependence of the equations.}
	\label{FigAppBFig2}
\end{figure}

We plot $f$, $g$, and $k$ for any point $\theta$ or $\phi$ on the convective element surface in 
Fig.\ref{FigAppBFig2} which makes evident that the conditions $f\left( {t,\theta } \right) \ll k\left( 
{t,\theta } \right)$ and $g\left( {t,\theta } \right) \ll k\left( {t,\theta } \right)$ hold in the temporal 
regime of interest (as mathematically proved in Lemma 1)(\footnote{Note how this is the contrary of what is 
claimed in MB16, where the authors apparently find, for the same solar model and using same equations, a 
failure of the conditions in  Eqs.\eqref{VF1} and \eqref{VF2} that vice versa we find valid here. Motivated by 
this discrepancy, we have provided here all the numbers adopted in our computation \textit{explicitly}, so 
that these equations can be tested easily.}).

\section{Conceptual limits and practical implications of the SFCT}\label{SecLimitsOfValidity}
\subsection{On the functional timescales of the SFCT}
The solution of the convective energy transport is achieved by the SFCT thanks to the uniqueness theorem 
(P14, Sec.6.2) which is a theorem that holds inside a \textit{linear}-response theory framework. Hence the 
theory inherits the limits of this theorem, and these implications of this limitation are better understood 
with an example.
The theorem states that, to the leading order, there is an invariant manifold solution of the system 
of equations of energy transfer inside a star. If an invariant solution exists, we might be tempted to get the 
invariance regime directly from the quintic Eq.\eqref{Quintic} by taking to infinity the temporal limit of the 
coefficients Eq.\eqref{QuinticCoeff}, thus obtaining after simple manipulation:
\[\mathop {\lim }\limits_{\tau  \to \infty } {\left. {\sum\limits_{i = 0}^5 {{c_i}{{\bar v}^i}} } 
\right|_{\chi  \sim {\tau ^2}}} = \frac{1}{9}{\bar v^2}\left( {\frac{{8\alpha {g_4}({\nabla _{{\rm{ad}}}} - 
{\nabla _{{\rm{rad}}}})}}{{{\nabla _{{\rm{ad}}}}}} + 9{{\bar v}^3}} \right) = 0.\]
Apparently, in this way we reduce the quintic to a cubic (as in the MLT case) whose only real solution
  for a convective region  would be
$$\bar v = \frac{{2\sqrt[3]{{\alpha {g_4}\left( {{\nabla _{{\text{rad}}}} - {\nabla _{{\text{ad}}}}} 
\right)}}}}{{{3^{2/3}}\sqrt[3]{{{\nabla _{{\text{ad}}}}}}}}.$$
\noindent
This is also confirmed  by a numerical investigation of the coefficient in Eq.\eqref{QuinticCoeff}: 
taking $\sum\limits_{i = 0}^5 {\frac{{{c_i}{\bar v^i}}}{{{\bar v^2}}}}  = 0$ and neglecting $\left\{ 
{{c_4},{c_5}} \right\} \ll \left\{ {{c_1},{c_2},{c_3}} \right\}$ we can obtain a solution for $\bar v$ in the 
resulting cubic that is completely indistinguishable from the $\bar v$ obtained form the quintic.

Unfortunately, despite its apparent simplicity, this way of using the uniqueness theorem is wrong because it 
is outside its founding hypotheses.
As explained in P14, the ``limit $\tau \to \infty$'' is a mathematical idealization for the dimensionless 
independent variable ($\tau$ in our case) to reach the asymptotic behavior of the solution, say for $t> 
t^{\infty}$. In the real situation of interest, $\tau$ cannot exceed a specific limit value securing that (i) 
the stellar region under consideration is still convective; (ii) the convective elements are still moving 
inside the convective areas; and (iii) the star itself does still exist.
The  state variables $p = p\left( {{\bm{x}};t} \right)$, density $\rho  = \rho \left( {{\bm{x}};t} \right)$, 
and temperature $T = T\left( {{\bm{x}};t} \right)$, set the real physical limit ``$\tau  \to \infty$'' (or 
``$\tau  \to \partial L$'' in the example of Sec.\ref{SecEssence}, $\tau $ normalized time), so that the limit 
of Eq.\eqref{Alpha}:
\[\mathop {\lim }\limits_{\tau  \to \infty } \alpha  = \mathop {\lim }\limits_{\tau  \to \infty } 
\frac{{ac{T^3}}}{{\kappa {\rho ^2}{c_P}}},\]
is coherently accounted for. However, the star at an infinite time does not exist any longer, and it is 
simple 
to prove that the modelling scheme for the SFCT flattens the state variable gradients to adiabatic 
stratifications: $\mathop {\lim }\limits_{\tau  \to \infty } \nabla  = \mathop {\lim }\limits_{\tau  \to 
\infty } {\nabla _e} = {\nabla _{{\text{ad}}}}$ and the size of the convective element would diverge $\mathop 
{\lim }\limits_{\tau  \to \infty } \chi  = \infty$ as expected (note indeed that one of the primary effects of 
the convection is to homogenize a stellar gradient).

To conclude, there is non mathematical shortcut to the quintic in  Eq.\eqref{Quintic} because of its 
physical implications.

\subsection{On the effect of the turbulence}
One of the known present limitations of the SFCT is the missing treatment of the turbulence.
We know that turbulence is one of the critical physical phenomena to consider, and we assume it to affect 
the SFCT both in the onset of the convection, e.g., in Fig.\ref{FigTimescale01}, \ref{FigTimescale02}, and 
\ref{FigTimescale03} for $t<{{t}^{(0)}}$, and at its asymptotic regime, say for $t>{{t}^{(\infty) }}$ in 
Fig.\ref{FigTimescale01}, \ref{FigTimescale02}, and \ref{FigTimescale03}. Nevertheless, for problems of 
interest to us as the generation of stellar tracks, the turbulence is well known to be irrelevant: results as 
from, e.g., \citet{2005AJ....130.1693B}, clearly show how stellar tracks based on turbulence-free theory such 
as the MLT are well capable of reproducing stellar isochrones that fit globular cluster stars from the main 
sequence up to the red-giant branch. MLT does not include any correction due to turbulence. Results as the 
ones shown in \citet{2005AJ....130.1693B} can indeed well be claimed as some of the most reliable proof of the 
validity of the MLT framework, and in turn, on the second-order relevance of the turbulent convection in 
reproducing true stellar observations.

Here, we can try to understand the role of the exchange of energy between the mean convective flow and 
the turbulence in the regime $t>{t}^{{(\infty) }}$. The regime $t<t_{{}}^{(0)}$, introduced in 
Sec.\ref{SecHomoSFCT} as a limiting case for the SFCT in a homogeneous isotropic medium, is hence not of 
interest.

If we assume that the turbulent cascade distribution function of the stellar plasma in ${{S}_{0}}$ is given 
by ${{f}_{0}}={{f}_{0}}\left( \bm{x},\bm{v};t \right)$ then, the mean convective flow is just the first order 
moment of this distribution ${{\bm{\bar{v}}}_{0}}=\int_{{}}^{{}}{{{\bm{v}}_{0}}{{f}_{0}}\left( \bm{x},\bm{v};t 
\right){{d}^{3}}\bm{x}}$, while the second-order centered moment is $\bm{\sigma 
}_{0}^{2}=\int_{{}}^{{}}{{{\left( {{\bm{v}}_{0}}-{{{\bm{\bar{v}}}}_{0}} \right)}^{\otimes 
2}}{{f}_{0}}{{d}^{3}}\bm{x}}$ (with ${{*}^{\otimes n}}$ the ordinary tensor product of order $n$). Because we 
are interested only in the case of the asymptotic regime $t>{{t}^{(\infty) }}$, we simplify the previous as 
$\int_{{}}^{{}}{{{\bm{v}}_{0}}{{f}_{0}}\left( \bm{x},\bm{v};t>{{t}^{(\infty) }} 
\right){{d}^{3}}\bm{x}}=\bm{v}_{0}^{\infty }$ and similarly for $\bm{\sigma }_{0}^{\infty }$ (we omit the 
square which is present here only for a historical reason while clearly $\bm{\sigma }$ can have negative cross 
terms.) It is simple to prove that the relation between the first and second moments is $\bm{\sigma 
}_{0}^{2}=\overline{\bm{v}_{0}^{\otimes 2}}-{{\bm{\bar{v}}}_{0}}{{\bm{\bar{v}}}_{0}}$.
If we time-average the Navier-Stokes equations (then usually referred as Reynolds-averaged Navier-Stokes), 
i.e., we get:
\begin{equation}
\left\langle \bm{v}_{0}^{\infty },{{\nabla }_{\bm{x}}} \right\rangle \bm{v}_{0}^{\infty }+
\overline{\left\langle {{\bm{v}}_{0}}-\bm{v}_{0}^{\infty },{{\nabla }_{\bm{x}}} \right\rangle \left( 
{{\bm{v}}_{0}}-\bm{v}_{0}^{\infty } \right)}={{\nabla }_{\bm{x}}}\frac{{{p}^{\infty }}}{{{\rho }^{\infty 
}}}	
\end{equation}
Moreover, noticing that the contribution from $\left\langle \left\langle \bm{v}_{0}^{\infty },{
{\nabla }_{\bm{x}}} \right\rangle ,{{\bm{v}}_{0}}-\bm{v}_{0}^{\infty } \right\rangle $ and $\left\langle 
{{\bm{v}}_{0}}-\bm{v}_{0}^{\infty },{{\nabla }_{\bm{x}}} \right\rangle \bm{v}_{0}^{\infty }$  cancel out when 
time-averaged, we arrive at the version of the Reynolds-averaged Navier-Stokes suited to our purpose:
\begin{equation}\label{EqNew}
\begin{gathered}
\left\langle {{\bm{v}}_0^\infty ,{\nabla _{\bm{x}}}} \right\rangle {\bm{v}}_0^\infty  =  - 
{\nabla _{\bm{x}}}\frac{{{p^\infty }}}{{{\rho ^\infty }}} + \left\langle {{\nabla _{\bm{x}}}, - \overline 
{\left( {{{\bm{v}}_0} - {\bm{v}}_0^\infty } \right)\left( {{{\bm{v}}_0} - {\bm{v}}_0^\infty } \right)} } 
\right\rangle  \\
=  - {\nabla _{\bm{x}}}\frac{{{p^\infty }}}{{{\rho ^\infty }}} - \left\langle {{\nabla _{\bm{x}}},
{{\bm{\sigma }}_0}} \right\rangle . \\
\end{gathered}
\end{equation}
While this equation holds within any small layer $L$, ${{p}^{\infty }}=\text{cnst.}$ 
and ${{\rho }^{\infty }}=\text{cnst.}$ , pressure and density retain their spatial dependence within the 
convective zone $V$ supposedly composed of several layers. At different $L$, and within $V$, we have that 
${{p}^{\infty }}={{p}^{\infty }}\left( \bm{x} \right)$ and $\rho ={{\rho }^{\infty }}\left( \bm{x} \right)$. 
Because of the results in Fig.\ref{FigTimescale01}, \ref{FigTimescale02}, and \ref{FigTimescale03} the 
temporal dependence can be safely omitted as  the structure of the star is unchanged over the time interval 
$\Delta t\simeq {{t}^{\infty }}$ of interest here. Eq.\eqref{EqNew} couples the mean flow ${{\bm{v}}^{\infty 
}}$, to the dispersion velocity $\bm{\sigma }_{0}^{\infty }$, i.e., to the statistical index of the turbulent 
motions of the stellar plasma. The components of $\bm{\tau }\equiv \rho {{\bm{\sigma }}_{0}}$ are indeed 
traditionally called the “Reynolds stresses” and have been the subject of extensive study related to the 
problem of the closure of the moment equations of ${{f}_{0}}$ \citep[e.g.,][]{2000tufl.book.....P, 
2002cstt.book.....L}.

For $t>{{t}^{\infty }}$, the ${{i}^{th}}$ component of the force acting on a surface element due 
to $\bm{\sigma }_{0}^{\infty }$ is written, e.g., as ${{\rho }^{\infty }}\bm{\sigma }_{0}^{\infty 
}{{d}^{2}}S$, so that for $t>{{t}^{\infty }}$ the rate of work ${{\dot{W}^{\infty}}}$ due to turbulent motions 
follows as ${{\rho }^{\infty }}\left\langle \bm{v}_{0}^{\infty },\bm{\sigma }_{0}^{\infty }{{d}^{2}}S 
\right\rangle$ for the whole convective shell of  volume $V$ bounded by the surface $S$ and we can write:
\begin{equation}\label{EqNew3}
\frac{d}{dt}{{W}^{\infty}}=\oint\limits_{S}{\left\langle \bm{v}_{0}^{\infty },{{\rho }^{\infty }}
\bm{\sigma }_{0}^{\infty }{{d}^{2}}S \right\rangle }=\int\limits_{V}{{{\rho }^{\infty }}\left\langle {{\nabla 
}_{\bm{x}}},\bm{v}_{0}^{\infty }\bm{\sigma }_{0}^{\infty } \right\rangle {{d}^{3}}V},
\end{equation}
where on the last equality we used Gauss's theorem. Hence, for a unit volume, we have that
\begin{equation}\label{EqNew4}
\begin{aligned}
\frac{d}{{{d^3}V}}\frac{d}{{dt}}{W^{\infty} } &= \left\langle {{\nabla _{\bm{x}}},{\bm{v}}_0^\infty 
{\bm{\sigma }}_0^\infty } \right\rangle  \\
&= {\rho ^\infty }\left\langle {{\nabla _{\bm{x}}},{\bm{\sigma }}_0^\infty } \right\rangle 
{\bm{v}}_0^\infty  + {\rho ^\infty }{\bm{\sigma }}_0^\infty \left\langle {{\nabla _{\bm{x}}},{\bm{v}}_0^\infty 
} \right\rangle,
\end{aligned}
\end{equation}
thus identifying the meaning of ${{\rho }^{\infty }}\left\langle {{\nabla }_{\bm{x}}},\bm{\sigma }_{0}^{\infty 
}\right\rangle$ also as a force. Because we defined $\bm{\sigma }_{0}^{\infty }$ just from an average 
procedure on the (unknown) distribution function, $\bm{\sigma }_{0}^{\infty }$ cannot create or destroy 
mechanical energy. Instead, $\bm{\sigma }_{0}^{\infty }$ can only represent the rate of change of kinetic 
energy per unit of volume.
For simplicity of argument, we exclude over/under-shooting of the convective elements so that we can safely 
assume we can null the convective flux at the boundary of the convective zone $V$, thus that the divergence 
$\left\langle {{\nabla }_{\bm{x}}},\bm{v}_{0}^{\infty }\bm{\sigma }_{0}^{\infty } \right\rangle$ statistically 
cancels out and the previous Eq.\eqref{EqNew4} balances the LHS with the RHS as follows:
\begin{equation}\label{EqNew5}
\int_{{}}^{{}}{-{{\rho }^{\infty }}\left\langle {{\nabla }_{\bm{x}}},\bm{\sigma }_{0}^{\infty } \right\rangle 
\bm{v}_{0}^{\infty }dV}=\int_{{}}^{{}}{\bm{\sigma }_{0}^{\infty }{{\rho }^{\infty }}\left\langle {{\nabla 
}_{\bm{x}}},\bm{v}_{0}^{\infty } \right\rangle dV}.
\end{equation}
Hence, if ${{\rho }^{\infty }}\left\langle {{\nabla }_{\bm{x}}},\bm{\sigma }_{0}^{\infty } \right\rangle$ 
is the force due to $\bm{\sigma }_{0}^{\infty }$ (per unit volume acting) on the mean flow, then ${{\rho 
}^{\infty }}\left\langle {{\nabla }_{\bm{x}}},\bm{\sigma }_{0}^{\infty } \right\rangle \bm{v}_{0}^{\infty }$ 
is the rate of work of this force, and $-{{\rho }^{\infty }}\left\langle {{\nabla }_{\bm{x}}},\bm{\sigma 
}_{0}^{\infty } \right\rangle \bm{v}_{0}^{\infty }$ the rate of loss of kinetic energy from the mean flow as a 
result of the turbulence that must equal precisely the rate of gain of the kinetic energy by the turbulence. 
Under the light of these considerations, it is evident from Eq.\eqref{EqNew5}, that $\bm{\sigma }_{0}^{\infty 
}$  gives rise to a net force acting on the mean flow whose rate of work is negative meaning that the mean 
flow loses energy to the turbulence, i.e., energy is transferred from the mean flow to the turbulence.

Concluding, we have proved that the mean flow is overestimated by the SFCT that, at the present stage, 
can provide only an upper limit to the convective flux (even if in a better way than what MLT does because it 
is a parameter-free theory).

It remains beyond the goal of the present paper to upgrade the SFCT to account for the turbulence cascade. 
In the light of the results in P16, it is probably not necessary for stellar tracks. Several approaches are 
available in the literature to account for turbulence parametrically, as well as the results of hydrodynamics 
3D simulations \citep[e.g.,][]{2015ApJ...809...30A} to which we refer the interested reader. See also 
\citet{2017LRCA....3....1K} for a recent review.

\section{Conclusions}\label{SecConclusion}

In this work, we investigated in some detail the key elements of the SFCT, in particular the pressure treatment. We have:
\begin{itemize}
	\item presented the pressure treatment of the SFCT in the limiting case of homogeneous isotropic 
intrastellar plasma and compared it with the MLT;
	\item presented a numerical validation of the SFCT equations with the simplest solar model available 
in 
the literature that complements the analytical treatment of P14;
	\item obtained the first numerical temporal estimations of the onset of the convection in a 
convective 
layer thus evidencing for the first time (numerically) a temporal dependence of classical Schwarzschild or 
Ledoux criteria that was not previously studied;
	\item remarked on the limitation of the SFCT regarding temporal integration and the treatment of turbulence.
\end{itemize}

Finally, in light of the above considerations, it is instructive to consider why the MLT captures to a good
 approximation the essence of the transport of energy inside a star. This happens because MLT contains a free 
parameter whose meaning is not  to merely quantify  the distance travelled  by ``an average'' eddy along the 
vertical direction inside a star, but also to take somehow into account  the whole phenomenon of the 
turbulence cascade of energy, the transmission of the energy to the flux, the interaction between eddies, and 
the complex pressure information transmission from a place to another, etc.

The MLT is not a hydrodynamic theory because no time evolution is considered in it. Therefore the question how
 far a convective element can travel has no simple physical meaning because there is no such a simple 
connection between the average motion of the turbulent eddies and the pressure scale length: the pressure 
scale length is a natural scale length of a star that is related to the dynamics of the convective elements 
through Eq.\eqref{Eq03} whereas the mixing-length, in reality, has little to do, if nothing at all, with a 
length or a radial motion.  As a matter of fact  the most successful models of turbulence do not actually deal 
with spherical bodies but rather with tube-like objects, even though the classic ``view'' of turbulence 
relates 
the transport of energy from large to small scales with a mechanism that passes from convective elements as 
blobs, to rolled up vortex sheets in a sequence of azimuthal vorticity, and to poloidal motions that sweep 
angular momentum outward radially to form sheets \citep[e.g.,][]{1972JFM....56..447K}.

The simple exercises we have described here cannot adequately capture the whole physical complexity of 
magnetic 
and turbulent convection at work in the real star. This is in part captured by the study of P14. Nevertheless, 
 the discussion presented here is valuable in clarifying the role played by the essential ingredients of the 
stellar convection, at least to the order required to reproduce the observed HR-diagrams.

The success of the SFCT is evident in the quality of the stellar models it generates (see P16) which nicely 
fit the positions of stars in the Hersprung-Russel $\left\{ \lg {T_{eff}},\lg {L/L_{\odot}} \right\}$ diagrams 
by continuously adapting itself to the ever-changing properties of a star during its evolutionary history and 
without requiring an external calibration on standard stars like the Sun.
Moreover, the mathematical exploitation of the uniqueness theorem in P14 can open new possibilities not only 
to eliminate of parameters whose nature and physical meaning are far from being clear (e.g., the 
mixing-length) but also to investigate closure relations of the hierarchy of hydrodynamic equations 
\citep[e.g.,][]{2002cstt.book.....L}.

The internal structure of stars can also be probed with other techniques such as deep helio-seismology and 
giro-seismology studies. Nevertheless, the SFCT constitutes a step forward with respect to the classical MLT 
with which it shares the significant merits of simplicity and easy usage, but which surpasses by eliminating 
the so-called mixing length parameter.  The SFCT  does not intend to replace or compete with more 
sophisticated theories of convection and 3D numerical simulations that, however,  appear to be still far from 
being readily incorporable into large stellar model datasets.

\section*{Acknowledgements}
We thank J. Kollmeier and E. K. Grebel for a careful reading of an early
version of this manuscript and for many constructive suggestions.
This work benefited from support by the National Science Foundation under Grant No. PHY-1430152 (JINA 
Centre for the Evolution of the Elements) and NASA grant No. NNX14AF84G.

\bibliographystyle{elsarticle_harv}
\bibliography{Pasetto_NA_2019}

\onecolumn

\appendix
\section{Non-inertial linear response theory for convective elements in spherical 
coordinates}\label{AppendixA}

We summarize here some results of the non-inertial response theory for
non-degenerate non-relativistic gas instabilities in spherical coordinates
developed by \citet{2015A&A...573A..48P}. This theory provides the context from which the SFCT is derived 
as a particular case. Our summary here is meant to clarify and better explain some of the assumptions made by  
P14 and also to highlight some of the considerations made in the present paper, in particular those about the 
fate of convective elements.

We limit ourselves to consider the convective elements as  blobs of constant density
${\rho _e}$ , slightly different from that of the stellar medium, with density $\rho$, at any generic 
location 
${\bm{r}}$ inside a spherical star centered in the inertial reference frame ${S_0}$.  The ideal surface 
(spherical) enclosing and separating the convective element from the rest of the medium is indicated 
with ${\mathbb{S}^2}:\Sigma \left( {\xi ,\theta ,\phi ;t} \right)$ and is implicitly defined by an equation of 
the type:
\begin{equation}\label{EqA1}
\Sigma \left( {\xi ,\theta ,\phi ;t} \right) \equiv \xi  -
\left( {{\xi _e}\left( t \right) + \eta \left( t \right)Y_l^m\left( {\theta
		,\phi } \right)} \right),
\end{equation}
where $Y_l^m$ are the spherical harmonics \citep[e.g.,][]{
	1965PhT....18l..70L} for $\theta  \in \left[ {0,\pi } \right[$ , $\phi  \in
\left[ {0,2\pi } \right[$ and  $\eta \left( t \right) \ll {\xi _e}\left( t
\right)\forall t$ is the radial perturbation of the unperturbed radius ${\xi=\xi _e}$ of
the convective element.
For this summary presentation, we will make use of the non-spherical perturbation of Eq.\eqref{EqA1} and 
highlight the various topics under examination, however without proving the results which can be instead found 
in Appendix A of \citet{2015A&A...573A..48P}.

Because a generic convective element forms from a density perturbation of the surrounding pre-existing 
stellar 
medium, there is undoubtedly continuity between the medium inside and outside the convective element at the 
surface $\Sigma$, i.e., the stellar medium satisfies the continuity equations on both  sides of the  surface 
$\Sigma$:
\begin{equation}\label{EqA2}
\left\{ \begin{gathered}
\frac{{\partial \Sigma }}{{\partial t}} + \left\langle
{{\nabla _{\bm{\xi }}}{\varphi _{e,{{\bm{v}}_1}}},{\nabla
		_{\bm{\xi }}}\Sigma } \right\rangle  = 0 \hfill \\
\frac{{\partial \Sigma }}{{\partial t}} + \left\langle
{{\nabla _{\bm{\xi }}}{\varphi _{{{\bm{v}}_1}}},{\nabla
		_{\bm{\xi }}}\Sigma } \right\rangle  = 0, \hfill \\
\end{gathered}  \right.
\end{equation}
where these conditions on the potential-flows ${\varphi _{e,{{\bm{v}}_1}}}$  inside
the convective element, and in the stellar medium ${\varphi
	_{{{\bm{v}}_1}}}$ are approximated to the leading order by:

\begin{equation}\label{EqA3}
\begin{gathered}
\varphi _{e,{{\bm{v}}_1}}^{} \cong {\xi ^l}Y_l^m\left(
{\frac{{\dot \eta }}{l}\xi _e^{l - 1} + \frac{{2\eta }}{l}{{\dot \xi }_e}\xi
	_e^l} \right) - \frac{{\xi _e^2{{\dot \xi }_e}}}{\xi }, \hfill \\
{\varphi _{{{\bm{v}}_1}}} \cong  - v\cos \left( \theta
\right)\xi \left( {1 + \frac{1}{2}\frac{{\xi _e^3}}{{{\xi ^3}}}} \right) -
\frac{{\xi _e^2{{\dot \xi }_e}}}{\xi } - \frac{{3\xi _e^{l + 1}}}{{\left( {l
			+ 1} \right){\xi ^{l + 1}}}}\\
\,\,\,\,\,\,\, \,\,\,\,  \eta \left( {\frac{v}{2}\frac{{\partial
			Y_l^m}}{{\partial \theta }}\sin \theta  + Y_l^m\left(
	{\frac{1}{3}\frac{{\dot \eta }}{\eta }{\xi _e} + \frac{2}{3}{{\dot \xi }_e}
		+ v\cos \theta } \right)} \right), \hfill \\
\end{gathered}
\end{equation}
respectively \citep[see Appendix A][]{2015A&A...573A..48P}. A convective element carries in its interior 
the stellar plasma which is at rest in ${S_1}$, i.e., no external flow is allowed to deeply penetrate inside 
the convective element (at the first order)  as long as the convective element retains its identity. 
Therefore, the difference between the two equations \eqref{EqA3} is limited to the velocity term present in 
the second equation and missing in the first one. It is easy  to prove that for $\eta  \to 0$ we recover the 
familiar potential flow approximation which is the basis of different studies, e.g., the pressure exerted by 
the intergalactic medium on dwarf galaxies \citep[see Eq.(6) in ][]{2012A&A...542A..17P}, the dynamics of the 
potential flow approximation for stellar convection in P14, etc.

At the surface radius $\left\| {\bm{\xi }} \right\| = {\xi
	_e} + \eta Y_l^m$   we impose  the condition of continuity for the stress vectors
${\bm{s}}$ in the inviscid approximation:
${\left\langle {{\bm{\hat
				n}},{\bm{s}}} \right\rangle _{\Sigma  = 0}} = {\left\langle
	{{\bm{\hat n}},{{\bm{s}}_e}} \right\rangle _{\Sigma  = 0}}.$
This can be calculated with the aid of Eq.\eqref{EqA3} and two energy conservation equations for the 
stellar medium \citep[see Eq.(7) of ][]{2012A&A...542A..17P}:

\begin{equation}\label{EqA4}
\frac{{\partial {\varphi _{{{\bm{v}}_1}}}}}{{\partial
		t}} + \frac{1}{2}\left\langle {{\nabla _{\bm{\xi }}}{\varphi
		_{{{\bm{v}}_1}}},{\nabla _{\bm{\xi }}}{\varphi _{{{\bm{v}}_1}}}}
\right\rangle  + \frac{p}{\rho } = f\left( t \right) - \left\langle
{{{\bm{a}}_{O'}},{\bm{\xi }}} \right\rangle
\end{equation}
and its analogues for ${\varphi _{e,{{\bm{v}}_1}}}$.  The time dependent
function $f$ is fixed by assuming hydrostatic equilibrium far away from $\Sigma $.

In P14 the treatment of the stellar convection is simplified by assuming (i) subsonic regime, (ii) hydrostatic
 equilibrium for the star and (iii) neglecting fluid treatment inside the convective element which is allowed 
merely to radiate from $\Sigma$.  While the first two hypotheses have been extensively commented on previous 
studies \citep[P14, ][and here]{ 2015A&A...573A..48P},  the third one can be easily understood as follows. A 
convective element density is only slightly different from that of the surrounding medium so that dissipative 
processes will act rapidly during its expansion. This implies that in such a case we can neglect self-gravity 
and ignore the Poisson equation for the matter contained in its interior. The loss of energy by the convective 
elements Eq.(60) of P14 is supposed to be only due to radiation from the surface where the radiative transfer 
equation is accounted. No mechanical dissolution of the convective elements is considered thus they keep their 
spherical shape until the end of their life. This approximation (apparently too restrictive for the purposes 
of \citet{2015A&A...573A..48P}) turned out to be  acceptable if one looks at the overall structure of stellar 
models calculated with the SFCT plus suitable boundary conditions \citep[e.g.,][]{2016MNRAS.459.3182P, 
Emanuela2016}.

Furthermore, by retaining the third hypotheses, we can rapidly derive an instability criterion that allows
 an understanding of the survival of the convective elements as they expand/contract moving upward/downward in 
an unstable convective layer as already mentioned in Section \ref{SecFate}. By developing the algebra in full 
\citep[see Eq.A30 in ][for more details]{2015A&A...573A..48P} we reach the criterion expressed by the quantity 
$\gamma^2 > 0$ (defined below) which allows us to investigate the onset of instabilities and the resulting 
dissolution of convective elements(\footnote{Traditionally the instability is indicated as ${\gamma ^2}$ and 
not as $\gamma$.  Although it can happen that ${\gamma ^2} < 0$,  the “square” is kept only for historical 
reasons and to save the connection with older studies.}). The quantity ${\gamma ^2}$ is defined as follows (RT 
and KH stand for Rayleigh-Taylor and Kelvin-Helmholtz respectively as introduced above):
\begin{equation}\label{EqA5}
\begin{gathered}
{\gamma ^2} = \gamma _{{\text{RT}}}^2 + \gamma _{{\text{a -
			RT}}}^2 + \gamma _{{\text{mix}}}^2 + \gamma _{\text{KH}}^2 + \gamma _I^2 \\
\left\{ \begin{gathered}
\gamma _{{\text{RT}}}^2 \equiv \frac{{3{A_ + }\left( {3{A_ +
			}{v^\parallel } + 8{{\dot \xi }_e}} \right)}}{{{{\left( {4{\xi _e}}
				\right)}^2}}}{v^\parallel } \\
\gamma _{{\text{a-RT}}}^2 \equiv \frac{{a_{O'}^\parallel
}}{2}\frac{{\left( {4lA - {A_ + }} \right)}}{{2{\xi _e}}} \\
\gamma _{{\text{mix}}}^2 \equiv \frac{9}{4}\frac{{2{A_{ -  -
		}}{A_ + }{F_1}}}{{{{\left( {2{\xi _e}} \right)}^2}}}{v^\parallel }{v^ \bot }
\\
\gamma _{\text{KH}}^2 \equiv \frac{9}{4}\frac{{{v^{ \bot 2}}\left(
		{A - 2\left( {{l_ - } + {F_2}} \right) + F_1^2} \right) + A}}{{4{{\left(
				{2{\xi _e}} \right)}^2}}} \\
\gamma _I^2 \equiv \frac{1}{2}\left( {\frac{{3\dot \xi
			_e^2}}{{2\xi _e^2}} + \frac{{A\left( {1 + 2l} \right){{\ddot \xi
				}_e}}}{{{\xi _e}}}} \right), \\
\end{gathered}  \right. \\
\end{gathered}
\end{equation}

\noindent
where the quantity $A$ is the generalized Atwood’s number
$$A
\equiv \frac{{{l_ + }{l_{ +  + }}{\rho _e} - {l_ - }l\rho }}{{{l_ + }{\rho
			_e} + l\rho }}, $$

\noindent
and $F_1$ and $F_2$ are two special function defined in Appendix B of \citet{2015A&A...573A..48P}:

$${F_1} \equiv \frac{1}{{Y_l^m}}\frac{{\partial
		Y_l^m}}{{\partial \theta }} \qquad {\rm and} \qquad
{F_2} \equiv
\frac{1}{{Y_l^m}}\frac{{{\partial ^2}Y_l^m}}{{\partial {\theta ^2}}}.$$

\noindent
The other quantities are
${l_ + } \equiv l + 1$ ,
${l_{ +  + }} \equiv l + 2$,
${A_ - }\equiv A - 1$ ,
$A_{--} \equiv  A-2$, etc.
and ${v^\parallel } \equiv v\cos \theta  \wedge {v^ \bot } \equiv v\sin
\theta $.
The nature and the role of the single terms in Eq.\eqref{EqA5} are reviewed in Section 
\ref{SecFate} and  \citet{2015A&A...573A..48P} where an extended analysis is made, and peculiar limits are 
treated. Here we limit ourselves to remark on  the asymmetric role that RT and KH instabilities have in the 
building up of ${\gamma ^2}$ in that while to the leading order the RT instability retains a 
dependence on the acceleration and velocity, i.e., both $\gamma _{{\text{RT}}}^2$  and $\gamma _{{\text{a - 
RT}}}^2$ survive the linearization of the equations, the KH instability contribution $\gamma _{\text{KH}}^2$, 
depends only on the velocity because its dependence on the acceleration is of the order of $O\left( {{\eta 
^2}} \right)$.

\end{document}